\documentclass[12pt, preprint]{aastex}

\shorttitle{Grain Growth in T Tauri Binaries}
\shortauthors{Skemer et al.}

\begin{document}

\title{Dust Grain Evolution in Spatially Resolved T Tauri Binaries\footnote{The observations reported here were obtained at the MMT Observatory, a facility operated jointly by the Smithsonian Institution and the University of Arizona.}}

\author{Andrew J. I. Skemer$^{1}$, Laird M. Close$^{1}$, Thomas P. Greene$^{2}$, Philip M. Hinz$^{1}$, William F. Hoffmann$^{1}$ and Jared R. Males$^{1}$}
\affil{$^{1}$Steward Observatory, Department of Astronomy, University of Arizona, 933 N. Cherry Ave, Tucson, AZ 85721}
\affil{$^{2}$NASA Ames Research Center, Moffett Field, CA 94035}

\begin{abstract}
Core-accretion planet formation begins in protoplanetary disks with the growth of small, ISM dust grains into larger particles.  The progress of grain growth, which can be quantified using 10$\micron$ silicate spectroscopy, has broad implications for the final products of planet formation.  Previous studies have attempted to correlate stellar and disk properties with the 10$\micron$ silicate feature in an effort to determine which stars are efficient at grain growth.  Thus far there does not appear to be a dominant correlated parameter.  In this paper, we use spatially resolved adaptive optics spectroscopy of 9 T Tauri binaries as tight as 0.25" to determine if basic properties shared between binary stars, such as age, composition, and formation history, have an effect on dust grain evolution.  We find with 90-95\% confidence that the silicate feature equivalent widths of binaries are more similar than those of randomly paired single stars, implying that shared properties do play an important role in dust grain evolution.  At lower statistical significance, we find with 82\% confidence that the secondary has a more prominent silicate emission feature (i.e., smaller grains) than the primary.  If confirmed by larger surveys, this would imply that spectral type and/or binarity are important factors in dust grain evolution.
\end{abstract}

\section{Introduction\label{Binaries Introduction}}

The core-accretion model for planet formation \citep[e.g.,][]{2007prpl.conf..591L} is a multi-step process beginning with the agglomeration of small interstellar medium (ISM) dust grains into larger particles.  Eventually, these particles reach the size of planetesimals and gravitationally attract each other and their surrounding gas.  Planet formation must occur quickly, as several physical processes are able to disperse the gas and dust of the protoplanetary disk.  Consequently, the timescale over which each aspect of core-accretion occurs is critical to the final outcome of the planet formation process.  The first step of core-accretion, where small ISM dust grains agglomerate into larger particles, is of particular importance as it sets the conditions for all of the other core-accretion steps.

Dust grain properties can be studied at a variety of wavelengths and locations in the disk \citep{2007prpl.conf..767N}.  Visible and near-infrared scattered light imaging can be used to study the grain sizes of submicron and micron surface dust at large radial distances.  Mid-infrared spectroscopy of the 10$\micron$ and 20$\micron$ silicate emission features can be used to determine the size and composition of submicron and micron sized surface dust at small radial distances.  Far-infrared and millimeter continuum observations probe deep into the disk mid-plane over the full radius of the disk and can be used to determine the size distribution and total mass of mm-cm dust grains.  Although these observations probe different grain sizes at different locations in the disk, it appears that they are somewhat correlated.  In particular, the equivalent width of the 10$\micron$ silicate feature is correlated with the slope of the submillimeter spectral energy distribution (SED), implying that dust grain growth proceeds concurrently through its various phases \citep{2010A&A...515A..77L}.  Additionally, the shape of the 10$\micron$ silicate feature is correlated with its amplitude, likely due to the simultaneous development of large amorphous grains and crystalline grains \citep{2003A&A...400L..21V,2003A&A...412L..43P,2005Sci...310..834A}.

Our current understanding of dust grain growth is that it begins during the Class II phase of star/planet formation, once the dust is collected into a circumstellar disk \citep{1991ApJ...381..250B,1994A&A...291..943O,1994A&A...288..929K}.  At this point, dust grains are expected to coagulate and settle rapidly \citep[$\sim10^{3}-10^{6}$ years depending on model assumptions;][]{2005AA...434..971D,2008AA...480..859B}.  However, Class II objects over a wide span of ages ($\sim$0.5-10 Gyr) have a diverse set of dust grain properties, implying that dust evolution is not simply a function of age \citep{2001A&A...365..476M,2003A&A...412L..43P,2005A&A...437..189V,2005Sci...310..834A,2006ApJ...639..275K,2009ApJ...703.1964F,2011arXiv1104.3574O}.  Even within individual clusters, which are approximately coeval, there is no apparent correlation between mid-infrared dust properties and stellar mass, luminosity, stellar accretion or disk mass \citep{2007ApJ...659.1637S,Watson2009}.  Over a large range of masses, taken between stars in different clusters with different ages, there appears to be a weak correlation between mass and silicate feature strength \citep{2006ApJ...639..275K,2009ApJ...696..143P}, which might be the result of another spectral-type dependent factor, such as X-rays or the location of the silicate emission zone \citep{2009A&A...508..247G,2007ApJ...659..680K}.  However, the general diversity of dust properties between similar stars of similar ages suggests that additional properties are needed to explain the evolution of the dust.

One way to isolate which stellar properties are important for dust grain evolution is to observe the dust grain properties of coeval young binaries.  Binaries share certain properties, such as age, composition and formation history, which are difficult to ascertain in individual stars.  If these properties are important for dust grain evolution, then we expect the dust to be similar in binary pairs.  However, if these properties are unimportant, then the binary stars will have a random distribution of dust characteristics, similar to those observed in single stars.  Additionally, if the shared properties are important, they might be hiding correlations between other stellar parameters and dust grain characteristics, which can be uncovered by taking advantage of the coevality of binaries.

In this paper \citep[and including the results of ][]{Skemer2010}, we present spatially resolved 10$\micron$ silicate spectroscopy of 8 Taurus-Aurigae\footnote{Taurus-Auriga is a low-mass star forming region with an age of $\sim$1-2 Myr and a distance of $\sim$140 pc \citep{1995ApJS..101..117K,1994AJ....108.1872K}.} binaries, which triples the sample size of spatially/spectrally resolved binaries in a single cluster at mid-infrared wavelengths.  Small diameter space-based telescopes are unable to perform high spatial resolution ($\lesssim 2-3$") observations in the mid-infrared ($>$8$\micron$) due to their large diffraction limits (additionally, the Infrared Spectrograph, IRS, on \textit{Spitzer} is no longer operational, so mid-infrared spectroscopy is not currently accessible from space).  Ground-based observatories have much larger diameters but suffer from high sky background, variable transmission and seeing, which explains why constructing even a small sample has been difficult.  We describe our observations and reductions in Section 2 and our corrections for extinction and equivalent width measurements in Section 3.  In Section 4, we perform statistical tests in an attempt to determine which properties affect dust grain-growth.  In particular, we ask the questions (1) \textit{Is grain growth correlated between binary pairs?} and (2) \textit{Does removing this effect reveal correlations between grain growth and other properties?}  We discuss the implications of our results and our conclusions in Section 5.

\section{Observations and Reductions \label{Binaries Observations Section}}
We observed 7 T Tauri binaries with MMTAO/BLINC-MIRAC4 over several nights in 2009 and 2010 (the object names, dates, and exposure times can be found in Table 1).  All of the objects were observed with the 6.5 meter MMT Telescope and its unique deformable-secondary adaptive optics system \citep[MMTAO;][]{2000PASP..112..264L, 2003SPIE.5169...17W, 2004SPIE.5490...23B}, which produces stable, high-Strehl (up to $\sim$98\%) point-spread-functions (PSFs) in the mid-infrared \citep{2003ApJ...598L..35C}.  We used the combination mid-IR nulling-interferometer/imager/spectrograph, BLINC-MIRAC4 (Bracewell Infrared Nulling Cryostat; Mid-IR Array Camera, Gen. 4), in its imager/spectrograph configuration \citep{1998SPIE.3354..647H,2000SPIE.4006..349H,Skemer09}.  

For each binary, we observed a bright spectroscopic standard, that doubled as a PSF star, before and after the binary observation.  We took acquisition images at 8.7$\micron$ to put the standard or binary in MIRAC4's 1"$\times$15" slit.  We then inserted the KRS-5 grism, which is in a filter wheel, near a pupil plane.  For the binaries, we azimuthally aligned both stars with the slit, either based on our 8.7$\micron$ images or on optical acquisition images for the AO-system's wavefront sensor.  The telescope's Cassegrain de-rotator was turned on to keep both stars in the slit.  We chopped perpendicularly to the slit with an 8" throw, and nodded along the slit (so that the binary was visible in both nods) with $\sim$3-5" nods, depending on the binary separation.  

The adaptive optics (AO) system is able to run at full-speed (550 Hz loop-speed) for stars brighter than V$\sim$12 mag, and sometimes fainter in good conditions.  Since some of our targets were fainter than this cutoff, we ran the AO system slower (100-150 Hz loop-speed) on these binaries, and their PSF star/spectroscopic standards.  Our PSF/spectroscopic standard stars were brighter at visible wavelengths than our target binaries, so we used neutral density filters in the AO camera to roughly match the wavefront sensor counts and limit differences in photon-noise error.

Our reduction methods are described in \citet{Skemer2010}.  One difference is that in this paper, instead of having an overall ``measurement error", we often measure relative errors (between binary sources) separately from absolute errors (between one or both binary components and a spectroscopic standard).  One benefit of this is that for sources with large ($>$2) mid-IR flux ratios, we can do an absolute calibration on just the brighter component with the best subset of our observations (i.e., close in airmass and time to bracketing spectroscopic standards and in photometric conditions), and build up S/N on the fainter companion in other conditions, without having to frequently switch between the object and a standard.  The method is also useful for tight binaries, where we can use a subset of the data to do the absolute calibration on the combined binary, and then use the full dataset to split the binary.  We use this method for all but GG Tau, and XZ Tau (2009 dataset), which were observed only briefly, so that all of the data were used for both relative and absolute calibrations.  The exposure times for our absolute calibrations (a subset of our full data) and our relative calibrations (our full data) are listed in Table 1.

We used the mid-IR standard $\epsilon$ Tau for all of our binaries except GG Tau, for which we used HR 1684 \citep{1999AJ....117.1864C}.  In all cases except GG Tau and XZ Tau (2009 dataset), we were able to measure telluric calibration errors \citep[as in][]{Skemer2010} with bracketing standards.  In the cases of GG Tau and XZ Tau (2009 dataset), which were both taken in photometric conditions with well-matched single spectoscopic standards in both airmass and time, we conservatively assume 3\% local errors and 5\% global flux errors outside of telluric ozone.

Most of the binaries in our survey have a large enough separation that splitting them is trivial with the 6.5 meter MMT and its adaptive optics system.  However, a few of our binaries (GG Tau, GN Tau, XZ Tau, which are separated by 0.25", 0.41" and 0.29" respectively) are near or below the diffraction limit ($\lambda/D=$0.32" at 10$\micron$) of the 6.5 meter MMT.  Because of MMTAO's high-Strehl and stability in the mid-infrared, it has resolved binaries as tight as 0.12" at 10$\micron$ using superresolution techniques \citep{2008ApJ...676.1082S}.  By using our spectroscopic standard as a 1-D PSF, we fit a double (binary) PSF, for each wavelength of our spectrum image, fixing the separation of the binary based on imaging data.  This gives us a measurement of the flux of each component relative to our spectroscopic standard.

For our tightest object, GG Tau, we also fixed the curvature of the grism trace to that of the PSF star's grism trace, which is legitimate as long as the object and PSF star are observed at a similar airmass to avoid the effects of atmospheric dispersion \citep{Skemer09}.  Both GG Tau and its PSF star were observed at 1.04 airmasses.

For GN Tau, the A and B components are of similar brightness in the optical, and because of their small separation, the AO system's wavefront sensor was somewhat less effective in the direction of the binary.  As a result, the binaries' PSF was slightly wider in the direction of the binary than the spectroscopic standard's PSF.  We found that convolving our spectroscopic standard's PSF with a 0.05"-$\sigma$ Gaussian greatly improved the residuals of the binary fit, although this smoothing did not significantly affect the measurement values of the fit.

\begin{figure}
\begin{center}
\includegraphics[angle=0,width=5.0in]{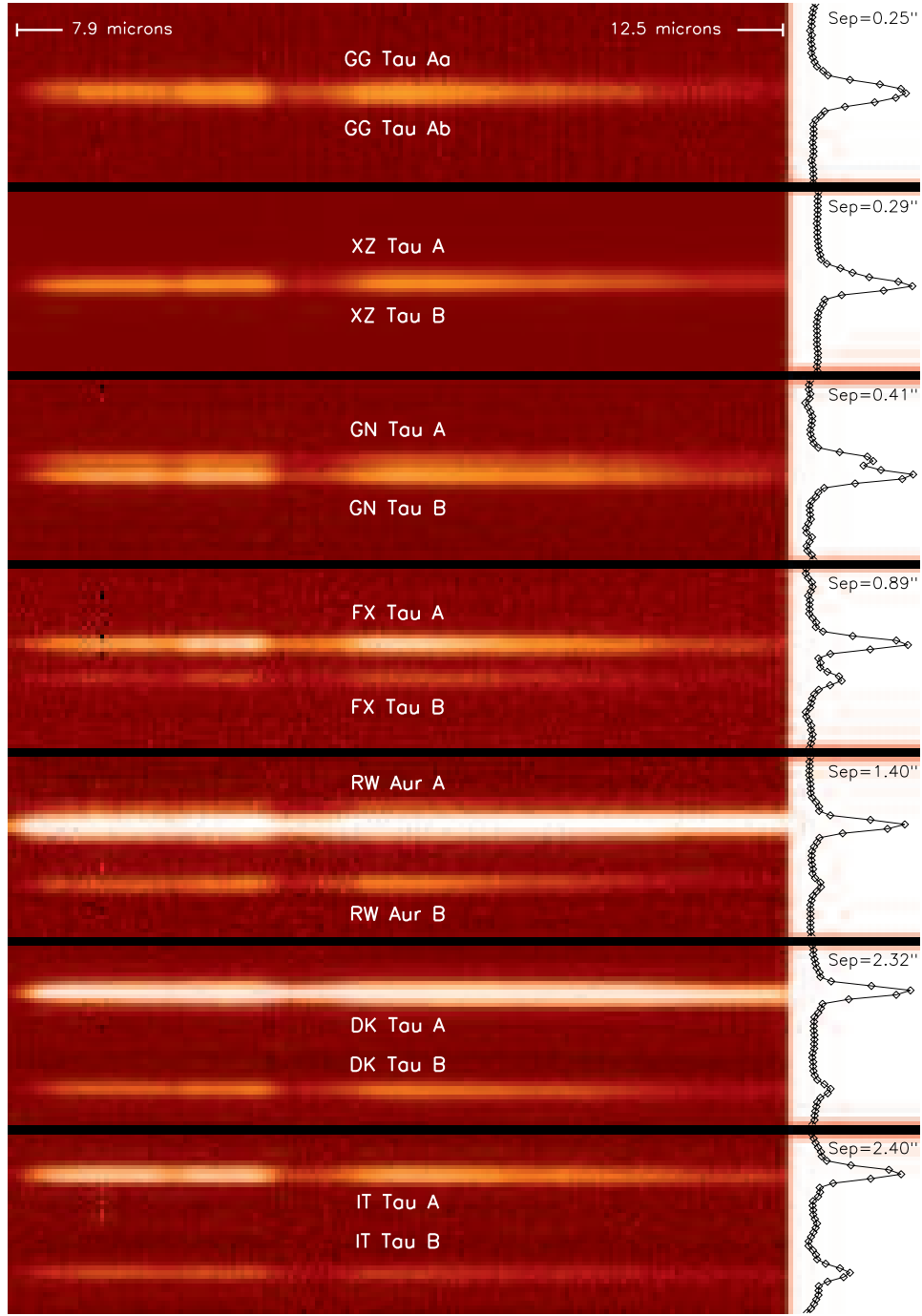}
\caption[Spectrally dispersed images of 7 T Tauri binaries]{Spectrally dispersed images of 7 T Tauri binaries that we have spatially and spectrally resolved for the first time at N-band.  The wavelength range (shown at the top) is 7.9$\micron$ to 12.5$\micron$ and has a suppressed flux at 9.7$\micron$ due to telluric ozone.  The images are each 4.4" in the spatial direction and are displayed with (different) linear stretches.  At the right of every spectrally dispersed image is a vertical (spatial direction) profile cut. GG Tau and XZ Tau are separated by less than the diffraction-limit of the 6.5 meter MMT but are superresolved due to the excellent stability of the MMT's adaptive optics PSF in the mid-infrared.  This is evident in the vertical profiles, which show that that GG Tau and XZ Tau are wider than the PSF (seen in the well-separated binaries below).
\label{Binaries spectra images}}
\end{center}
\end{figure}

Spectrally dispersed images for each of our seven binaries are shown in Figure \ref{Binaries spectra images}.  Each frame shows both components of a binary, with the primary on top and the secondary on the bottom (primary being defined as the brighter component in the optical).  A vertical cut through each image is shown to the right.  Our tightest binaries, GG Tau Aa-Ab and XZ Tau A-B are separated by less than the MMT's $\sim$0.32" diffraction limit.  However, the vertical cut through the spectrally dispersed images shows a wider spectrum, and with knowledge of the binary separation (obtained from acquisition images) and our high-Strehl ($\sim$98\%) PSF, we are able to spatially resolve the spectra.

Spectroscopy for each of our seven binaries (plus a second epoch on one of the binaries, XZ Tau), is presented in Figure \ref{Binaries spectra}.  In all cases, the A component is to the left and the B component is to the right.  The error bars that coincide with the Earth's 9.7 $\micron$ ozone feature reflect the larger calibration uncertainties of our data at these wavelengths.  

A summary of all (our survey plus literature) spatially and N-band spectrally resolved binaries/multiples in the Taurus-Auriga star forming region is presented in Table 2.  We rebin published spectra from other instruments (DD Tau A-B and FV Tau A-B from \citealp{Honda2006} and GI-GK Tau from \citealp{Watson2009}) to the wavelength sampling used in our reductions in order to keep our results consistent.  The uncertainties for DD Tau A-B and FV Tau A-B are from \citet{Honda2006}.  The uncertainties for GI-GK Tau are dominated by our binning method \citep[see][]{Skemer2010} because of \textit{Spitzer}'s high sensitivity.  Flux loss due to \textit{Spitzer} slit mispointings become irrelevant in our later analysis of silicate equivalent widths.

\section{Analysis}
\subsection{Correction for Extinction\label{Binaries De-extinction}}
Our mid-IR spectra must be corrected for extinction to avoid contamination by the silicate feature of the ISM extinction curve \citep{1985ApJ...288..618R}.  In Taurus, where most Class II objects are extincted by $A_{V}<$3 \citep{1995ApJS..101..117K}, the effect is not as great as in other star forming regions.  However, some of our objects have larger extinctions \citep[the largest of which is FV Tau AB, with $A_{V}=5.3$;][]{White_Ghez_2001}.  As shown in Figure 1 of \citet{2009ApJ...703.1964F}, these extinctions will have a strong influence on the resulting silicate emission feature if left uncorrected.

We correct for extinction in our measured mid-IR spectra using $A_{V}$ values from \citet{2009ApJ...703.1964F} and references within (listed in Table \ref{Binaries Measurements}), which will allow us to make a consistent comparison between our binaries and the single stars of \citet{2009ApJ...703.1964F} in Section 4.1.  We adopt the same extinction laws as \citet{2009ApJ...703.1964F}: \citet{1990ARA&A..28...37M} with $R_{V}=5$ for objects with $A_{V}<3$ and \citet{2009ApJ...693L..81M} for $A_{V}\ge3$.  Since our objects are binaries, we correct extinction in both components equally, as interstellar extinction is thought to be the same between tight binary components \citep{White_Ghez_2001,2003ApJ...583..334H}.  The one exception is for our widest binary, GI-GK Tau, where \citet{2009ApJ...703.1964F} cite separate extinction values for each component.

Figure \ref{deextincted spectra} shows our extinction corrected spectra.  There are several examples of binaries that have strikingly similar silicate features, both in terms of their strengths and their shapes.  Notable example include FX Tau, GN Tau, IT Tau, DD Tau, FV Tau and GI-GK Tau.  We quantitatively compare the strength of silicate features in binary pairs, using equivalent width measurements, in Section \ref{Binaries correlated?}.  More detailed comparisons of their detailed silicate properties would be useful for determining whether individual dust components (for example, certain crystalline species) are correlated between binary pairs, but that test is not addressed further in this paper.  

Our sample shows a diversity of silicate properties.  Strong features (implying small dust grains) are present in GI Tau, GK Tau, RW Aur B and GG Tau Ab, among others. Weak features (no small grains) appear in IT Tau A and B and most notably in XZ Tau B, which shows negligible emission in 2009.  XZ Tau, which was observed twice, shows strong variability, as was seen in multi-epoch \textit{Spitzer} spectroscopy \citep{2009ApJ...706L.168B}.  Our data indicate that silicate variability is present in both sources, and because of the strong variation we observe in the flux ratio of the binary, both silicate features likely affect the unresolved \textit{Spitzer} variability.

\subsection{Equivalent Width Measurements: A Signature of Grain Growth}\label{EqW Measurements section}
Numerous authors have noted the correlation between the shape and the strength of the 10 micron silicate feature, where a strong feature generally implies the presence of small grains.  Since the ISM is composed of small, amorphous silicate grains \citep{2004ApJ...609..826K}, it is assumed that circumstellar disks where these grains are observed have not undergone dust processing/grain growth, while disks with larger and/or more crystalline grains have begun grain growth.  The 10 $\micron$ silicate feature is only sensitive to a subsample of dust in the circumstellar disk: silicate dust grains between $\sim$0.1$\micron$ and $\sim$5$\micron$, that are at the optically thin surface of the disk and at the proper temperature to be in the silicate emission zone \citep{2007ApJ...659..680K}.  While the 10$\micron$ silicate feature is being used to probe grain properties in the upper-layers of the circumstellar disk, millimeter wavelengths can probe the cooler and deeper layers of the disk, which contain larger mm-cm sized grains.  The shape of the millimeter continuum slope can be used to probe a characteristic mid-plane grain size, which has also been used to infer grain-growth.  In a highly important work, \citet{2010A&A...515A..77L} demonstrated a correlation between the strength of the silicate feature (using equivalent width) and the slope of the millimeter continuum.  This means that despite the limited emission zone of the 10$\micron$ silicate feature, it does appear to be a marker for dust grain-growth throughout young circumstellar disks.  This result has practical importance, because there are many more measurements of 10$\micron$ silicate features than millimeter slopes, due to the decreasing brightness of these objects with wavelength.  Binarity, in particular, is best studied at shorter wavelengths where the system can be spatially resolved.

\citet{2010A&A...515A..77L}'s result correlates the equivalent width of the silicate feature with dust grain sizes in the disk midplane.  Similarly, silicate de-composition techniques (see \citealt{2009ApJS..182..477S,2009ApJ...695.1024J,2010A&A...520A..39O,2011arXiv1104.3574O} and a review by \citealt{2009ASPC..414...77W}) establish that the silicate feature equivalent width is connected to the size of dust grains in the upper layers of the disk.  Based on these two correlations, we use the equivalent width of the silicate feature as a proxy for grain-growth for the rest of this paper.  While silicate feature de-composition techniques still contain more information than this simple approach, equivalent width is a robust, single-number statistic, that is easily measured with ground-based systems that are limited in sensitivity, spectral range, and spectral resolution.

We calculate equivalent width by integrating $(F-F_{continuum})/F_{continuum}$, where the continuum is a linear fit of our flux measurements between 8.07-8.27$\micron$ and 12.56-12.95$\micron$.  Errors are calculated using the Monte Carlo approach described in \citet{Skemer2010}.  The dominant error source for most of our MMTAO/BLINC-MIRAC4 data is the spectroscopy at the edges of the silicate band, which define the (linear-fit) silicate continuum.  These regions are particularly faint from the ground because of the low atmospheric transmission $\le8\micron$ and $\ge12.5\micron$ combined with decreasing flux at the longer wavelengths.  The other difficult region in the N-band is the $9.7\micron$ telluric ozone feature, where our absolute calibrations tend to be much worse than elsewhere in the spectrum (see Figure \ref{Binaries spectra}).  For the purposes of calculating equivalent width, we interpolate over the ozone region.

The results of our equivalent width measurements of binaries, as well as 8.1 and 12.7 $\micron$ continuum photometry and colors are presented in Table \ref{Binaries Measurements}.  We include our MMTAO/BLINC-MIRAC4 measurements, as well as the binaries from \citet{Honda2006} and \citet{Watson2009}, which we interpolate in the same fashion as the MMTAO/BLINC-MIRAC4 for consistency.  We do not include T Tau, UY Aur or XZ Tau, which are not used in Section \ref{Binaries correlated?} because they are self-extincted and variable.  All of the quantities listed are calculated with the extinction corrected spectra.

Our ground-based equivalent width measurements use a smaller than desirable wavelength range ($\sim$8-12.5$\micron$) due to truncation by atmospheric transmission.  Space-based equivalent width measurements, motivated by the true width of the silicate feature, use larger wavelength ranges (for example, \citet{2010A&A...515A..77L} use 7.5-13.0$\micron$).  Additionally, space-based equivalent width measurements often use polynomial continuum fits instead of linear fits and do not interpolate over telluric ozone, as is done for our ground-based measurements.  For these reasons, it is worth checking that the equivalent widths calculated using our ground-based prescription are well-correlated with the equivalent widths calculated using typical space-based prescriptions.  We use a set of single stars in Taurus (described in Section \ref{Binaries correlated?}), whose extinction corrected equivalent widths are published by \citet{2009ApJ...703.1964F} based on \textit{Spitzer}/IRS spectroscopy.  We obtained the \textit{Spitzer} data, and applied our MIRAC4 binning procedure and equivalent width measurements prescription.  A comparison of our measurements with the measurements of \citet{2009ApJ...703.1964F} is shown in Figure \ref{Binaries EQW compare}.  The space-based prescription (true) equivalent widths are well-correlated with the ground-based prescription (truncated) equivalent widths, with the exception of one outlier (DM Tau, which has a large quadratic continuum term \citep{Watson2009} that is not accounted for in the ground-based prescription).  This implies that ground-based (truncated) equivalent widths are reliable proxies for the space-based (true) equivalent widths.

\section{Statistical Tests}
\subsection{Is Grain Growth in Binaries Correlated?\label{Binaries correlated?}}
Properties shared between binary stars, such as age, composition and formation history, might be linked to the growth of dust grains in young stars.  We test for this by determining if the silicate features of binary stars are more similar than the silicate features of randomly selected pairs of single stars.  We use the ratio of extinction corrected equivalent width measurements (Table \ref{Binaries Measurements}) to quantify the similarity of two silicate features, using the reciprocal of the ratio when the ratio is $<1$.

The binary stars are drawn from our sample of spatially and spectrally resolved N-band spectra in Taurus-Auriga (Table \ref{Binaries Binaries list}), where we exclude systems that contain infrared companions, that are thought to be self-extincted and variable \citep[T Tau, UY Aur, XZ Tau, as shown by][]{2010AA...517A..16V,Skemer2010,2009ApJ...706L.168B}.  This leaves us with 9 binaries for the test.  For the pairs of single stars, we use the extinction corrected equivalent widths of 26 Taurus-Aurigae single stars from \citet{2009ApJ...703.1964F}, where the single stars are listed in \citet{2006ApJS..165..568F}.  We exclude sources with unknown and high ($A_{\rm V}>8$) extinctions, and we avoid DG Tau, which is known to have a variable silicate feature \citep{2000AAS...197.4707W,2009ApJ...706L.168B} and HK Tau, which has a self-extincting disk \citep{1998ApJ...502L..65S}.  We recalculate the extinction-corrected equivalent widths of the single stars using our ground-based prescription (see Section \ref{EqW Measurements section}) and list these values in Table \ref{Single Measurements}.  The single stars and binaries are similarly distributed on an HR diagram (Figure \ref{Binaries HR}), allowing us to make a reasonably unbiased comparison.

We do an N=$10^{4}$ Monte Carlo simulation to randomly pair single stars and to randomly sample the error bars for our binary equivalent width ratios (a histogram of the simulation output is shown in Figure \ref{Binaries single_compare}.  In 90\% of our trials, the average (mean) of the equivalent width ratios of the 9 binaries is closer to 1 than the average equivalent width ratios of 9 randomly selected pairs of single stars.  Thus, with 90\% confidence, we find that silicate features in binary pairs are more similar than silicate features in randomly paired single stars, implying that grain growth is correlated in binaries.

There is one systematic effect that is likely to have increased the observed differences between the silicate features of the single star pairs.  The extinction measurements for each single star correspond to an approximately 15\% error in equivalent width \citep{2009ApJ...703.1964F}.  This will cause a residual differential extinction between the single star pairs, that should on average make the single star silicate features look more different than they actually are.  We add 15\% more uncertainty to the single star equivalent widths in our Monte Carlo simulation and find that our confidence is only changed by 2\%.  Thus, the effect of residual extinction in the single star pairs is small.  Similiarly the extinction corrections for the binaries, described in Section 3.1, assumes that the binaries share the same interstellar extinction.  If this assumption is incorrect there would be a residual differential extinction between the binary pairs, that should on average make the binary silicate features look more different than they actually are.  This bias is similarly small to the single star bias.

Our single star sample includes 3 transitions disks (GM Aur, DM Tau and LkCa 15), which have larger than normal silicate equivalent widths \citep[see Figure \ref{Binaries EQW compare} and][]{2009ApJ...703.1964F}.  We include them in our single star sample because some of our binaries have similarly large silicate equivalent widths (for example GG Tau Ab, which has a ``truncated" $EqW=3.33\pm0.40$).  Excluding the transition disks (while including GG Tau) would change our confidence that grain growth is correlated in binaries to 83\%.

Of our 9 binaries, GG Tau Aa-Ab has the largest disparity between its silicate feature equivalent widths.  GG Tau A is a 0.25" binary with circumstellar disks around each component and a spatially resolved circumbinary disk \citep{1994A&A...286..149D} that is unusually massive \citep{1996ApJ...458..312J}.  Some authors have reported tentative detections of a dust streamer onto GG Tau Ab \citep{1996ApJ...463..326R,2011AA...528A..81P}, which might be replenishing the small dust grains detected in our silicate spectroscopy \citep{Skemer2010}.  The presence of streamers has also been inferred in tighter binary systems via periodic accretion \citep{1997AJ....113.1841M,2007AJ....134..241J}.  If the dust steamer(s) replace the dust grains around each components at different rates, the silicate features would not be ``coeval" as we have assumed throughout this analysis.  Excluding GG Tau (and the transition disks) from our Monte Carlo simulation would change our confidence that grain growth is correlated in binaries to 95\%.

\subsection{Does the Coevality of Binaries Reveal a Correlation of Grain Growth and other Properties?\label{Binaries others?}}
Numerous authors have used single stars (and unresolved binaries) to determine if stellar and disk properties might correlate with grain-growth.  The strongest correlation is between X-ray luminosity and dust grain crystallinity and size \citep{2009A&A...508..247G,2009ApJ...701..571R}.  A weaker and possibly dependent correlation might exist between stellar mass and dust-grain size \citep{2006ApJ...639..275K,2009ApJ...696..143P}.  It is possible that age and/or formation history effects are obscuring the correlations between grain growth and stellar/disk properties.  Here, we use the coevality of binaries to effectively remove the age/formation history affects and test correlations between grain-growth and various other parameters.

As in Section \ref{Binaries correlated?}, we use the 9 binaries from our sample that do not have edge-on disks/self-extinction.  For each of these binaries, we have compiled a table of spatially resolved ancillary stellar properties (Table \ref{Binaries star properties}) and a table of infrared flux ratios for determining disk colors (Table \ref{Binaries disk properties}).  We plot 6 different stellar/disk properties versus binary equivalent width ratios in Figure \ref{Binaries correlations}.  With our small sample size of binaries (9), doing too many correlation tests would result in spurious correlation detections.  Thus, the properties we have chosen are intended to have the most plausible links to grain-growth.  Spectral type is an obvious catch-all, that for a fixed age, is related to the temperature, luminosity and mass of the star.  Bolometric magnitude relates to the total flux hitting the dust-grains, and should define the ``silicate emission zone" \citep{2007ApJ...659..680K}.  Stellar mass relates to the gravity-induced sinking of large dust grains to the disk midplane and disk dynamics in general.  $H_{\alpha}$ relates to accretion, which will produce UV photons that can anneal the dust.  K-L colors relate to the geometry of the disk's puffed-up inner-rim.  K-N colors relate to the disk geometry at the location of the silicate emission zone.  

For each frame in Figure \ref{Binaries correlations}, the x-axis shows a difference/ratio measurement for a star/disk property.  For example, frame (a) shows the difference in spectral type between the A component and the B component of the binary. In all frames, the y-axis shows the ratio of the equivalent width ratios.  Vertical dotted lines show equal spectral types, luminosities, masses, $H_{\alpha}$ equivalent widths, and colors.  Horizontal dotted lines show equal silicate equivalent widths.  This divides each frame into quadrants.  We do a coarse correlation analysis by determining how many binaries are in the top-right or bottom-left quadrants versus the top-left or bottom-right quadrants.  This allows us to easily answer questions of the form ``Do the stars with earlier spectral types have larger or smaller silicate feature equivalent width?"  Since we are dealing with coeval binary pairs, this directly relates to whether spectral type is correlated with grain-growth.

In 7 out of 9 binaries, the star with the earlier spectral type has a smaller silicate equivalent width than the star with the later spectral type.  Using the binomial distribution with $p=50\%$, the probability that 7 or more out of 9 binaries would have larger equivalent widths in either the earlier-type components or the later-type components is 18\%.  This probability is not low enough to rule out the null hypothesis that spectral type is not correlated with silicate feature equivalent widths (especially given that it is 1 of 6 tests shown in Figure \ref{Binaries correlations}).  No other parameter in Figure \ref{Binaries correlations} has a stronger correlation signal than spectral type.  Additionally, none of the parameters in Figure \ref{Binaries correlations} have an obvious visual trend.  Thus, we do not find evidence for any correlations with equivalent width, other than the marginal one observed for spectral type.

\section{Discussion and Conclusions}
Our mid-infrared adaptive optics survey has produced 8 new spatially resolved silicate spectra of young Taurus-Aurigae binaries \citep[including UY Aur, which was published in][]{Skemer2010}.  This effectively triples the number of spatially and spectrally resolved binaries in the Taurus-Auriga cluster, and provides the first opportunity to draw statistical conclusions about the spatially resolved silicate features of binaries in any young cluster.  Our fundamental questions are (1) \textit{Is grain growth correlated between binary pairs?} and (2) \textit{Does removing this effect reveal correlations between grain growth and other properties?}

We find with 90\% confidence (when including GG Tau) or 95\% confidence (when excluding GG Tau) that the silicate features of binaries are more similar than the silicate features of randomly paired single stars (our ambivalence in including GG Tau stems from its large circumbinary disk, which might be replenishing the dust in the inner disks via streamers).  The similarity of the silicate features in our binary pairs implies that one or more shared binary properties (such as age, composition or formation history) plays an important role in dust grain evolution.  

One possible explanation for our findings is that the environment in which a star forms affects its potential to grow dust grains, and eventually planets.  It is a fundamental prediction of core-accretion that giant planets form more efficiently in metal-rich systems, based on the increased solid material available for building planetary cores \citep{2004ApJ...604..388I,2004ApJ...616..567I}.  However, the core-accretion scenario is predicated by small dust grains coagulating quickly enough that aerodynamic drag does not remove too many meter-sized objects from the system \citep[a problem, which is commonly known as the ``meter-sized barrier";][]{1977MNRAS.180...57W}.  \citet{2008AA...480..859B} have shown that the removal of meter-sized objects due to radial drift can be overcome by increasing the disk's initial dust-to-gas ratio from 1\% to 2\%.  The result is a natural consequence of increasing the rate of grain growth by decreasing the time between collisions.  Since binary stars form from the same fragmented core material, it is possible that grain growth in binary stars will proceed at similar rates in each component due to their shared initial gas-to-dust ratio.

Binarity itself might also affect the silicate features of young stars.  Circumstellar disks can be dynamically perturbed by a companion star, which introduces spiral structure and truncates the radial extent of the disk \citep{1994ApJ...421..651A}.  In this configuration, large particles collide and fragment more frequently, which speeds up dust grain evolution \citep{1995Icar..114..237D} while decreasing the maximum particle size in the disk \citep{2011A&A...527A..10Z}.  Dust streamers from a wider circumbinary disk can also replenish solid material in one or both of the circumstellar disks \citep{1996ApJ...467L..77A}.  The dust streamers are likely to contain un-evolved, ISM-like dust grains that dominate the appearance of the silicate feature \citep{Skemer2010}.  There are two young Taurus-Aurigae binaries with directly imaged circumbinary disks: GG Tau \citep{1994A&A...286..149D} and UY Aur \citep{1996A&A...309..493D}.  In each case, we find that one of the components has an unusually prominent silicate feature, indicating the presence of small grains.  \citet{2008ApJ...673..477P} were not able to detect a difference between the silicate features of (unresolved) medium-separation binaries and single stars, which suggests that any effect must be subtle or uncommon.  However, spatially resolved studies of the silicate features in young binaries might reveal different dust-grain growth mechanisms.

In Section \ref{Binaries others?}, we found that seven of our nine binaries (i.e. 82\% confidence of a trend using the binomial distribution) have larger silicate equivalent widths (smaller grains) in the secondary than the primary.  This pattern was also observed in two binaries by \citet{2003A&A...412L..43P}.  Assuming this trend is confirmed by larger samples of spatially resolved silicate spectroscopy, the cause could be based on the dynamics of binaries or simply a spectral-type/mass/luminosity correlation with the speed of dust grain growth.  The former explanation could be tested by seeing if the correlation is connected with binary separation.  

Our study has been limited by its small sample size, due to the complexity and limitations of ground-based 10 $\micron$ spectroscopy and adaptive optics, which were necessary to build up the current sample.  Larger ground-based telescopes, such as the LBT and ELTs along with large space-based telescopes, such as JWST will be able to dramatically increase the sample of spatially resolved silicate spectra of young binaries.

\acknowledgements
The authors thank Elise Furlan, Mitsuhiko Honda and Dan Watson for providing us with their published spectra.  We also thank Daniel Apai, Ilaria Pascucci and Kevin Flaherty for useful discussions regarding grain-growth and silicate variability.  AJS acknowledges the NASA Graduate Student Research Program (GSRP) for its generous support of this project.  We also thank the MMT staff, especially Mike Alegria, Morag Hastie and Ricardo Ortiz for operating the AO system during a difficult set of observations.

\clearpage

\begin{deluxetable}{lcccccccccccc}
\tabletypesize{\scriptsize}
\tablecaption{MMTAO/BLINC-MIRAC4 Observations of T Tauri Binaries}
\tablewidth{0pt}
\tablehead{
\colhead{Binary} &
\colhead{Dates (UT)} &
\colhead{Relative Calibration} &
\colhead{Absolute Calibration} &
\\
\colhead{} &
\colhead{} &
\colhead{On-source Time (s)\tablenotemark{a}} &
\colhead{On-source Time (s)\tablenotemark{a}} &
}
\startdata
DK Tau & 2010 Jan 3 & 2190 & 1260 \\
FX Tau & 2010 Jan 2 and 6 & 1200 & 570 \\
GG Tau A & 2009 Oct 2 & 1200 & 1200 \\
GN Tau & 2010 Jan 2 & 2550 & 750 \\
IT Tau & 2010 Jan 3 and 4 & 2385 & 2385 \\
RW Aur & 2010 Jan 5 and 6 & 765 & 630 \\
XZ Tau & 2009 Jan 14 & 160 & 160 \\
XZ Tau & 2010 Jan 5 & 1230 & 270 \\
\enddata

\tablenotetext{a}{Relative fluxes between binary pairs were measured using all available data.  Absolute calibrations (using the brighter star or the ``un-resolved" binary pair) were measured using the best subset of data to minimize errors caused by changing atmospheric conditions.}

\label{Binaries All Observations}
\end{deluxetable}

\clearpage

\begin{deluxetable}{lcccccccccccc}
\tabletypesize{\scriptsize}
\tablecaption{N-band Spatially and Spectrally Resolved Taurus Binaries}
\tablewidth{0pt}
\tablehead{
\colhead{Binary} &
\colhead{Separation (")} &
\colhead{Separation Ref.} &
\colhead{N-band spectra Ref.} &
}
\startdata
DD Tau A-B                      & 0.56      & 1         & 2 \\
DK Tau A-B                      & 2.32      & 1         & this work \\
FV Tau A-B\tablenotemark{a}     & 0.71      & 1         & 2 \\
FX Tau A-B                      & 0.89      & 1         & this work \\
GG Tau Aa-Ab\tablenotemark{b}   & 0.25      & this work & this work \\
GI-GK Tau\tablenotemark{c}      & 12.9      & 3         & 4 \\
GN Tau A-B                      & 0.41      & this work & this work \\
%Haro 6-37 A-B\tablenotemark{d} & 2.62      & 1         & 2 \\
IT Tau A-B                      & 2.4       & 5         & this work \\
RW Aur A-B\tablenotemark{d}     & 1.40      & 1         & this work \\
T Tau N-Sa-Sb                   & 0.69/0.13 & 6         & 7 \\
UY Aur A-B                      & 0.88      & 1         & 8 \\
XZ Tau A-B\tablenotemark{e}     & 0.29      & this work & this work\\
\enddata

\tablecomments{A list of T Tauri binaries in the Taurus-Auriga star-forming regions with spatially resolved N-band spectroscopy.  We only include binaries separated by $<30"$, which are then likely to be bound and coeval \citep{2009ApJ...704..531K}.  Due to non-negligible orbital motion in the cases of GG Tau, GN Tau and XZ Tau, astrometry is measured from 8.7$\micron$ acquisition images preceding our N-band spectroscopy.  We use $\alpha$ Gemini as our astrometric standard.  T Tau Sa-Sb also has fast orbital motion, which is parameterized, most recently, by \citet{Kohler2008}}

\tablenotetext{a}{does not include FV Tau /c, which is a 0.7" binary located 12.3" from FV Tau \citep{White_Ghez_2001}}
\tablenotetext{b}{does not include GG Tau B, which is a 0.25" binary located 10.3" from GG Tau A \citep{White_Ghez_2001}}
\tablenotetext{c}{GK Tau is itself a 2.4" binary \citep{Hartigan_Strom_Strom_1994}.  However, GK Tau B was not detected in our 8.7 micron imaging.}
%\tablenotetext{d}{Haro 6-37 A is itself a 0.3" binary \citep{1999A&A...343..831D}, which in this case was not resolved.} 
\tablenotetext{d}{RW Aur was found to be a triple by \citep{1993AJ....106.2005G}, but \citep{White_Ghez_2001} claim the 3rd star was probably a false detection.}
\tablenotetext{e}{XZ Tau A is itself a 0.09" binary at mm wavelengths \citep{2009ApJ...693L..86C}.}

\tablerefs{
(1) \citet{White_Ghez_2001};
(2) \citet{Honda2006};
(3) \citet{Hartigan_Strom_Strom_1994};
(4) \citet{2006ApJS..165..568F}, \citet{Watson2009} and others;
(5) \citet{Duchene_Monin_1999};
(6) \citet{Kohler2008};
(7) \citet{Ratzka};
(8) \citet{Skemer2010}
}
\label{Binaries Binaries list}
\end{deluxetable}

\clearpage

\begin{deluxetable}{lcccccccccccc}
\tabletypesize{\scriptsize}
\tablecaption{Extinction Corrected Silicate/Continuum Measurements of T Tauri Binaries}
\tablewidth{0pt}
\tablehead{
\colhead{} &
\colhead{$A_{V} (mag)$\tablenotemark{a}} &
\colhead{Silicate EqW ($\micron$)} &
\colhead{8.1 $\micron$\tablenotemark{b} flux (Jy)} &
\colhead{12.7 $\micron$\tablenotemark{c} flux (Jy)}
}
\startdata
DD Tau A	  & 1.0 & 0.95$\pm$0.06 & 0.45$\pm$0.01 & 0.73$\pm$0.00 \\
DD Tau B	  & 1.0 & 1.53$\pm$0.11 & 0.34$\pm$0.01 & 0.43$\pm$0.01 \\
DD Tau A/B	  &     & 0.62$\pm$0.06 & 1.33$\pm$0.06 & 1.72$\pm$0.03 \\
\hline
DK Tau A	  & 1.3 & 2.80$\pm$0.18 & 0.66$\pm$0.03 & 0.78$\pm$0.02 \\
DK Tau B	  & 1.3 & 1.76$\pm$0.25 & 0.15$\pm$0.01 & 0.14$\pm$0.01 \\
DK Tau A/B	  &     & 1.60$\pm$0.18 & 4.51$\pm$0.14 & 5.52$\pm$0.30 \\
\hline
FV Tau A	  & 5.3 & 1.03$\pm$0.10 & 0.30$\pm$0.00 & 0.32$\pm$0.01 \\
FV Tau B	  & 5.3 & 1.44$\pm$0.11 & 0.43$\pm$0.01 & 0.54$\pm$0.01 \\
FV Tau A/B	  &     & 0.72$\pm$0.09 & 0.70$\pm$0.03 & 0.59$\pm$0.02 \\
\hline
FX Tau A	  & 2.0 & 2.61$\pm$0.39 & 0.14$\pm$0.01 & 0.17$\pm$0.02 \\
FX Tau B	  & 2.0 & 2.34$\pm$0.43 & 0.05$\pm$0.00 & 0.07$\pm$0.01 \\
FX Tau A/B	  &     & 1.12$\pm$0.26 & 2.91$\pm$0.22 & 2.45$\pm$0.35 \\
\hline
GG Tau Aa	  & 1.0 & 0.91$\pm$0.25 & 0.36$\pm$0.02 & 0.39$\pm$0.04 \\
GG Tau Ab	  & 1.0 & 3.33$\pm$0.40 & 0.24$\pm$0.02 & 0.28$\pm$0.02 \\
GG Tau Aa/Ab	  &     & 0.27$\pm$0.08 & 1.47$\pm$0.10 & 1.38$\pm$0.14 \\
\hline
GI Tau		  & 2.3 & 1.93$\pm$0.05 & 0.68$\pm$0.01 & 0.79$\pm$0.00 \\
GK Tau		  & 1.1 & 2.76$\pm$0.09 & 0.69$\pm$0.02 & 0.86$\pm$0.00 \\
GI Tau/GK Tau	  &     & 0.70$\pm$0.03 & 0.99$\pm$0.03 & 0.92$\pm$0.00 \\
\hline
GN Tau A	  & 3.5 & 1.04$\pm$0.21 & 0.14$\pm$0.01 & 0.15$\pm$0.01 \\
GN Tau B	  & 3.5 & 1.66$\pm$0.20 & 0.23$\pm$0.01 & 0.21$\pm$0.01 \\
GN Tau A/B	  &     & 0.62$\pm$0.12 & 0.59$\pm$0.03 & 0.75$\pm$0.04 \\
\hline
IT Tau A	  & 3.8 & 0.86$\pm$0.18 & 0.20$\pm$0.01 & 0.17$\pm$0.01 \\
IT Tau B	  & 3.8 & 1.06$\pm$0.37 & 0.08$\pm$0.00 & 0.07$\pm$0.01 \\
IT Tau A/B	  &     & 0.81$\pm$0.26 & 2.51$\pm$0.07 & 2.29$\pm$0.27 \\
\hline
RW Aur A	  & 0.5 & 1.10$\pm$0.07 & 1.05$\pm$0.05 & 1.26$\pm$0.05 \\
RW Aur B	  & 0.5 & 2.45$\pm$0.48 & 0.12$\pm$0.01 & 0.11$\pm$0.01 \\
RW Aur A/B	  &     & 0.45$\pm$0.10 & 8.43$\pm$0.62 & 11.81$\pm$1.59 \\
%\hline
%XZ Tau A (2009)   & 2.9 & 0.78$\pm$0.30 & 0.26$\pm$0.02 & 0.62$\pm$0.05 \\
%XZ Tau B (2009)   & 2.9 & 0.12$\pm$0.15 & 2.31$\pm$0.13 & 2.61$\pm$0.09 \\
%XZ Tau A/B (2009) &     & 3.78$\pm$4.29 & 0.11$\pm$0.01 & 0.24$\pm$0.02 \\
%\hline
%XZ Tau A (2010)   & 2.9 & 1.94$\pm$0.20 & 0.42$\pm$0.01 & 0.67$\pm$0.03 \\
%XZ Tau B (2010)   & 2.9 & 0.67$\pm$0.09 & 1.72$\pm$0.05 & 2.01$\pm$0.04 \\
%XZ Tau A/B (2010) &     & 2.88$\pm$0.49 & 0.24$\pm$0.00 & 0.33$\pm$0.02 \\

\enddata

\tablecomments{All error bars are calculated by taking the interquartile range of our Monte Carlo simulation, and dividing by 1.34.  This is a robust estimate of a Gaussian sigma.}

\tablenotetext{a}{$A_{V}$ values from \citet{2009ApJ...703.1964F}}
\tablenotetext{b}{Flux measured between 8.07$\micron$ and 8.27$\micron$}
\tablenotetext{c}{Flux measured between 12.56$\micron$ and 12.95$\micron$}

\label{Binaries Measurements}
\end{deluxetable}

\clearpage

\begin{deluxetable}{lcccccccccccc}
\tabletypesize{\scriptsize}
\tablecaption{Single Stars Used in Comparison with Binary Stars}
\tablewidth{0pt}
\tablehead{
\colhead{} &
\colhead{$A_{V} (mag)$\tablenotemark{a}} &
\colhead{Spectral Type\tablenotemark{a}} &
\colhead{Silicate EqW ($\micron$)\tablenotemark{b}}
}
\startdata
04108+2910         & 1.4 & M0   & 0.21 \\
AA Tau             & 1.8 & K7   & 1.18 \\
BP Tau             & 1.0 & K7   & 1.89 \\
CI Tau             & 2.0 & K7   & 1.61 \\
CW Tau             & 2.8 & K3   & 0.69 \\
CX Tau             & 1.3 & M0   & 1.05 \\
CY Tau             & 1.7 & K7   & 0.42 \\
DE Tau             & 1.2 & M0   & 1.21 \\
DL Tau             & 1.5 & K7   & 0.40 \\
DM Tau             & 0.7 & M1   & 1.43 \\
DN Tau             & 0.6 & M0   & 0.51 \\
DO Tau             & 2.0 & M0   & 0.65 \\
DP Tau             & 0.6 & M0.5 & 1.06 \\
DR Tau             & 1.2 & ...  & 0.80 \\
DS Tau             & 1.1 & K5   & 1.66 \\
F04147+2822        & 2.5 & M4   & 1.75 \\
FN Tau             & 1.4 & M5   & 1.25 \\
GM Aur             & 1.2 & K3   & 3.50 \\
GO Tau             & 2.0 & M0   & 1.49 \\
HP Tau             & 2.8 & K3   & 1.69 \\
IP Tau             & 0.5 & M0   & 2.63 \\
IQ Tau             & 1.4 & M0.5 & 1.26 \\
LkCa 15            & 1.2 & K5   & 3.98 \\
RY Tau             & 2.2 & G1   & 4.02 \\
V410 Anon 13       & 5.8 & M6   & 0.99 \\
V836 Tau           & 1.1 & K7   & 2.02 \\
\enddata

\tablenotetext{a}{$A_{V}$ values and spectral types from \citet{2009ApJ...703.1964F}}
\tablenotetext{b}{Equivalent Widths are based on \textit{Spitzer} spectra from \citet{2009ApJ...703.1964F} and are calculated using our ground-based prescription as described in Section \ref{Binaries De-extinction}}

\label{Single Measurements}
\end{deluxetable}

\clearpage

\begin{deluxetable}{lcccccccccccc}
\tabletypesize{\scriptsize}
\tablecaption{Ancillary Stellar Properties of Taurus Binaries}
\tablewidth{0pt}
\tablehead{
\colhead{Binary} &
\colhead{Spectral Type} &
\colhead{$H_{\alpha}$ EqW ($\AA$)} &
\colhead{Ref.} &
\colhead{$M_{bol}$ (mag)\tablenotemark{a}} &
\colhead{$M_{star}/M_{\Sun}$\tablenotemark{a}}

}
\startdata
DD Tau A   & M3.5                 & 206  & 1                  & 5.08 & $0.43^{+0.13}_{-0.23}$ \\
DD Tau B   & M3.5                 & 635  &                    & 5.40 & $0.41^{+0.11}_{-0.19}$ \\
\hline
DK Tau A   & K9                   & 31   & 2                  & 3.66 & $0.71^{+0.09}_{-0.06}$ \\
DK Tau B   & M1                   & 118  &                    & 5.35 & $0.61^{+0.06}_{-0.04}$ \\
\hline
FV Tau A   & K5                   & 15   & 3,1                &      &                        \\
FV Tau B   & K6\tablenotemark{b}  & 63   &                    &      &                        \\
\hline
FX Tau A   & M1                   & 13  & 4                  & 4.91 & $0.62^{+0.05}_{-0.03}$ \\
FX Tau B   & M4                   & 1.0  &                    & 5.97 & $0.28^{+0.16}_{-0.32}$ \\
\hline
GG Tau Aa  & M0                   & 57   & 1,3\tablenotemark{c} & 4.50 & $0.73^{+0.09}_{-0.08}$ \\
GG Tau Ab  & M2                   & 16   &                    & 4.87 & $0.64^{+0.03}_{-0.03}$ \\
\hline
GI Tau     & K6                   & 20   & 5,6                &      &                        \\
GK Tau     & K7                   & 22   &                    &      &                        \\
\hline
GN Tau A   & M2.5 (unresolved)\tablenotemark{d}    & 59 (unresolved) & 7 & & \\
GN Tau B   & & & & & \\
\hline
IT Tau A   & K3                   & 21.7 & 4                  & 3.84 & $1.0^{+0.3}_{-0.1}$    \\
IT Tau B   & M4                   & 147  &                    & 5.98 & $0.28^{+0.15}_{-0.32}$ \\
\hline
RW Aur A   & K1                   & 76   & 3                  & 3.11 & $1.4^{+0.6}_{-0.7}$    \\
RW Aur B   & K5                   & 43   &                    & 5.01 & $0.86^{+0.11}_{-0.10}$ \\
\enddata

\tablenotetext{a}{$M_{bol}$ and $M_{star}$ are calculated by \citet{2009ApJ...704..531K}}
\tablenotetext{b}{Note that \citet{2003ApJ...583..334H} found no photospheric absorption lines in their STIS spectrum of FV Tau B.}
\tablenotetext{c}{\citet{2003ApJ...583..334H}'s GG Tau spectrum is saturated at $H_{\alpha}$, so we use $H_{\alpha}$ equivalent width measurments from \citet{White_Ghez_2001}.}
\tablenotetext{d}{Although GN Tau does not have a published resolved spectral type, it has been resolved in the optical \citep{1996ApJ...469..890S}, near-infrared \citep{White_Ghez_2001} and mid-infrared (this work).  Throughout this range of wavelengths, it has a near equal flux ratio, so it is likely that the binary consists of two nearly equal mass stars.}

\tablerefs{
(1) \citet{2003ApJ...583..334H};
(2) \citet{1998A&A...339..113M};
(3) \citet{White_Ghez_2001};
(4) \citet{Duchene_Monin_1999};
(5) \citet{1995ApJS..101..117K};
(6) \citet{Hartigan_Strom_Strom_1994};
(7) \citet{2003ApJ...582.1109W}
}

\label{Binaries star properties}
\end{deluxetable}

\clearpage

\begin{deluxetable}{lcccccccccccc}
\tabletypesize{\scriptsize}
\tablecaption{Flux Ratios of Taurus Binaries}
\tablewidth{0pt}
\tablehead{
\colhead{Binary} &
\colhead{K Flux Ratio} &
\colhead{L Flux Ratio} &
\colhead{Near-IR Ref.} &
\colhead{12.7$\micron$ Flux Ratio}
}
\startdata
DD Tau A-B   & 1.17$\pm$0.06   & 1.72$\pm$0.13   & 1   & 1.72$\pm$0.03  \\
DK Tau A-B   & 4.21$\pm$0.30   & 4.02$\pm$0.13   & 1   & 5.52$\pm$0.30  \\
FV Tau A-B   & 1.49$\pm$0.01   & 0.69$\pm$0.02   & 1   & 0.59$\pm$0.02  \\
FX Tau A-B   & 2.20$\pm$0.02   & 2.97$\pm$0.03   & 1   & 2.45$\pm$0.35  \\
GG Tau Aa-Ab & 1.95$\pm$0.16   & 2.23$\pm$0.01   & 1   & 1.38$\pm$0.14  \\
GI-GK Tau\tablenotemark{a}    & 0.77$\pm$0.03   & 0.77$\pm$0.07  & 2,3 & 0.92$\pm$0.00  \\
GN Tau A-B   & 1.17$\pm$0.03   & 1.25$\pm$0.03   & 1   & 0.75$\pm$0.04  \\
IT Tau A-B   & 6.37$\pm$0.76   & 16.0$\pm$4.4    & 1   & 2.29$\pm$0.27  \\
RW Aur A-B   & 4.23$\pm$0.16   & 6.16$\pm$0.24   & 1   & 11.81$\pm$1.59 \\

\enddata

\tablenotetext{a}{Photometry for GI Tau and GK Tau have been extinction corrected individually using the $A_{V}$ values listed in Table \ref{Binaries Measurements}.}

\tablerefs{
(1) \citet{White_Ghez_2001};
(2) \citet{2006AJ....131.1163S};
(3) \citet{2006ApJ...647.1180L}
}

\label{Binaries disk properties}
\end{deluxetable}

\clearpage

\clearpage

\begin{figure}
\begin{center}
\includegraphics[angle=00,width=5.0in]{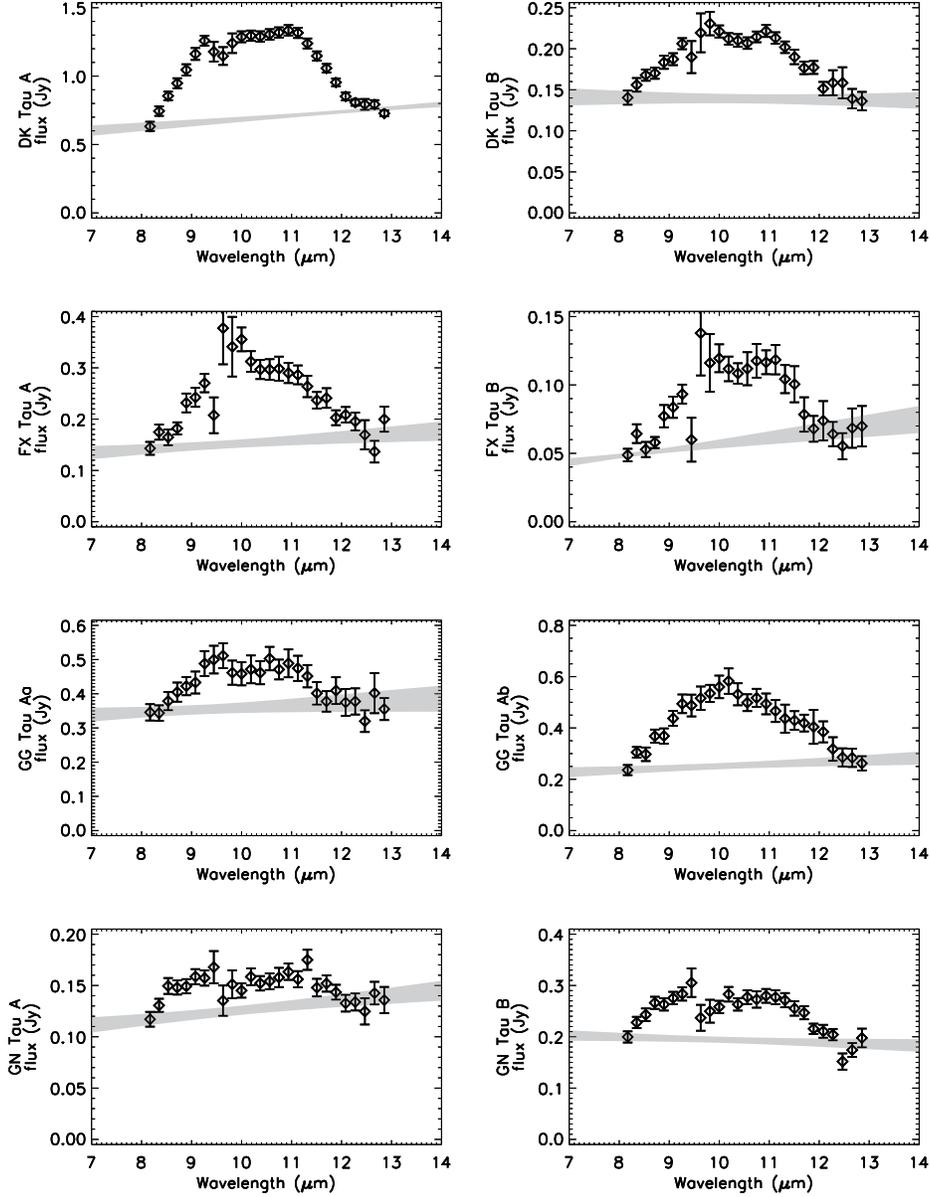}
\caption[Spectra of 7 T Tauri binaries]{Spectra of 7 T Tauri binaries that we have spatially and spectrally resolved for the first time at N-band.  XZ Tau is shown twice because we observed it once in 2009 and once in 2010.  In all cases, the primary is shown on the left and the secondary is shown on the right.  The grey region on each plot shows the 1$\sigma$ range of our continuum fit, which we use when calculating equivalent widths.
\textit{Continued}
\label{Binaries spectra}}
\end{center}
\end{figure}

\clearpage

\addtocounter{figure}{-1}

\begin{figure}
\begin{center}
\includegraphics[angle=00,width=5.0in]{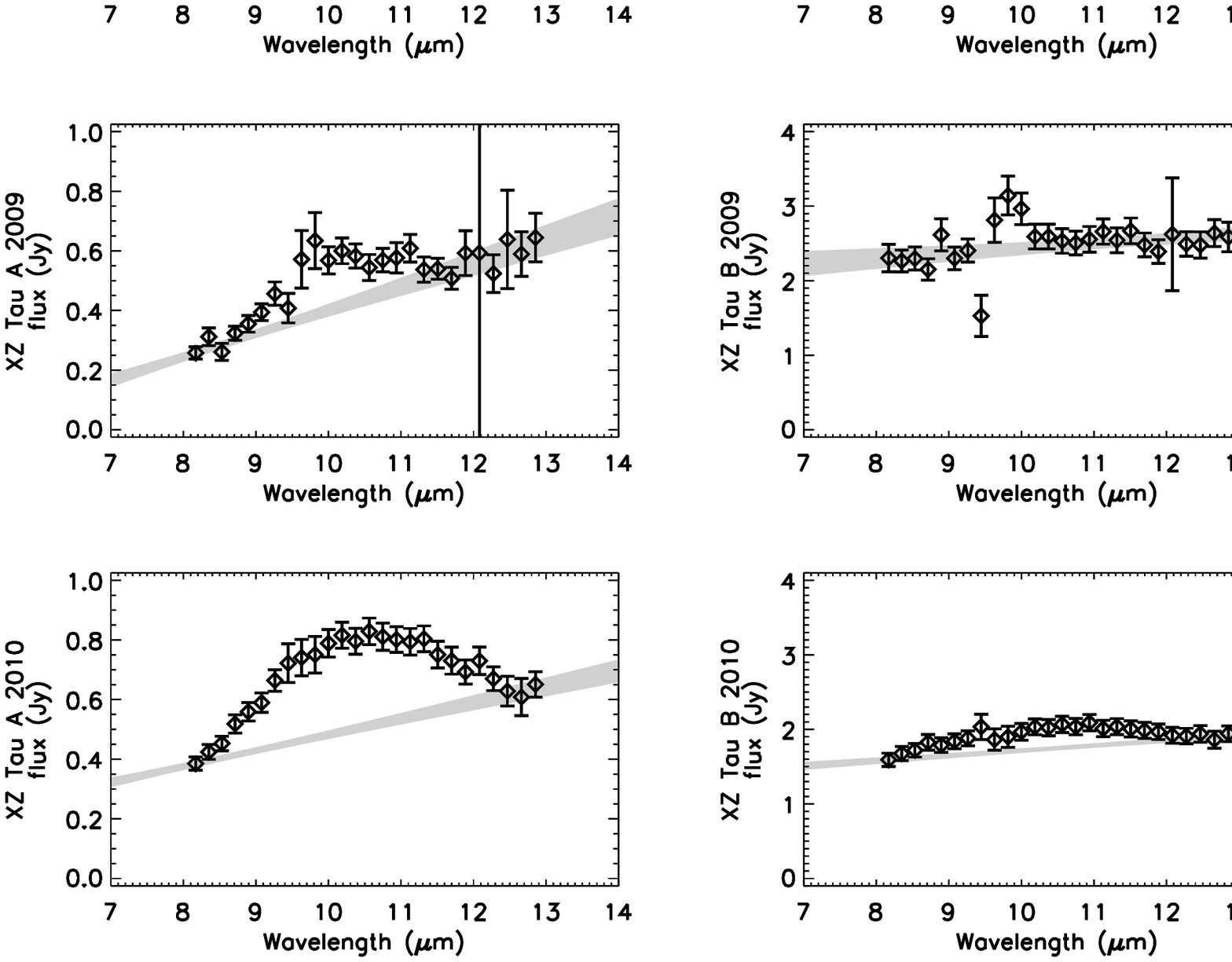}
\caption{\textit{Continued}}
\end{center}
\end{figure}

\clearpage

\begin{figure}
\begin{center}
\includegraphics[angle=00,width=5.0in]{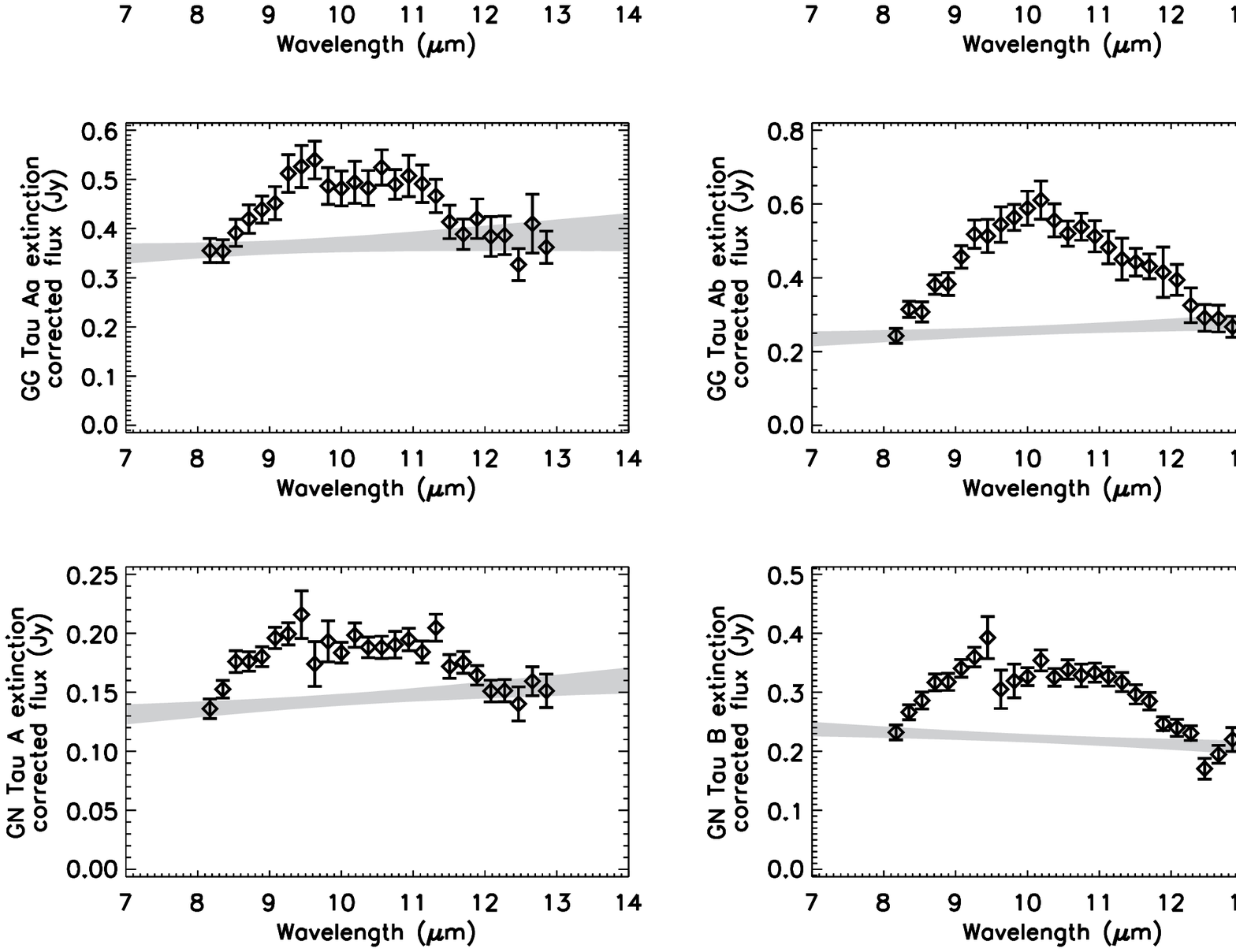}
\caption[Extinction Corrected Spectra]{Extinction corrected spectra from Figure \ref{Binaries spectra} with 3 additional binaries from \citet{Honda2006} and \citet{Watson2009}.
\textit{Continued}
\label{deextincted spectra}}
\end{center}
\end{figure}

\clearpage

\addtocounter{figure}{-1}

\begin{figure}
\begin{center}
\includegraphics[angle=00,width=5.0in]{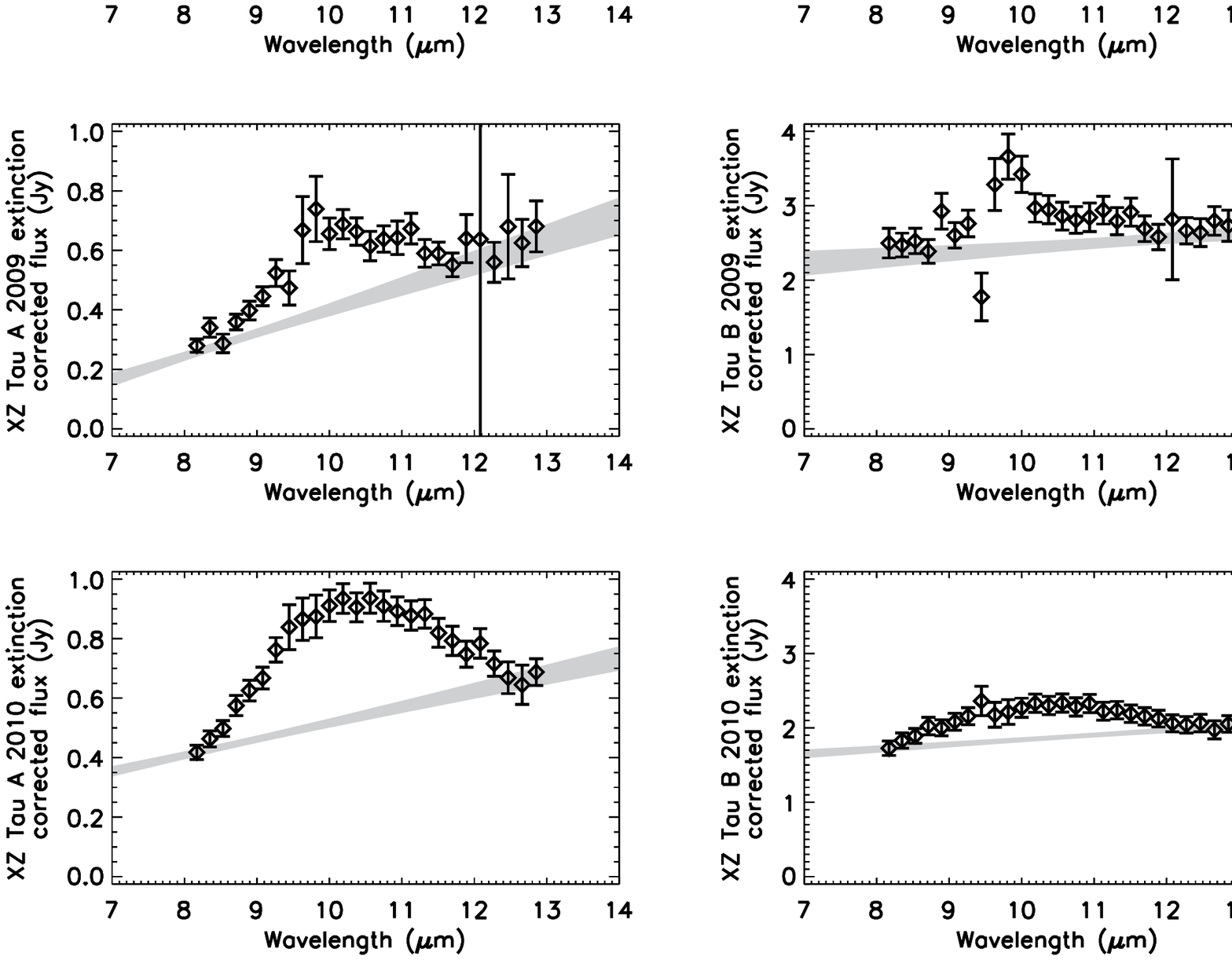}
\caption{\textit{Continued}}
\end{center}
\end{figure}

\clearpage

\addtocounter{figure}{-1}

\begin{figure}
\begin{center}
\includegraphics[angle=00,width=5.0in]{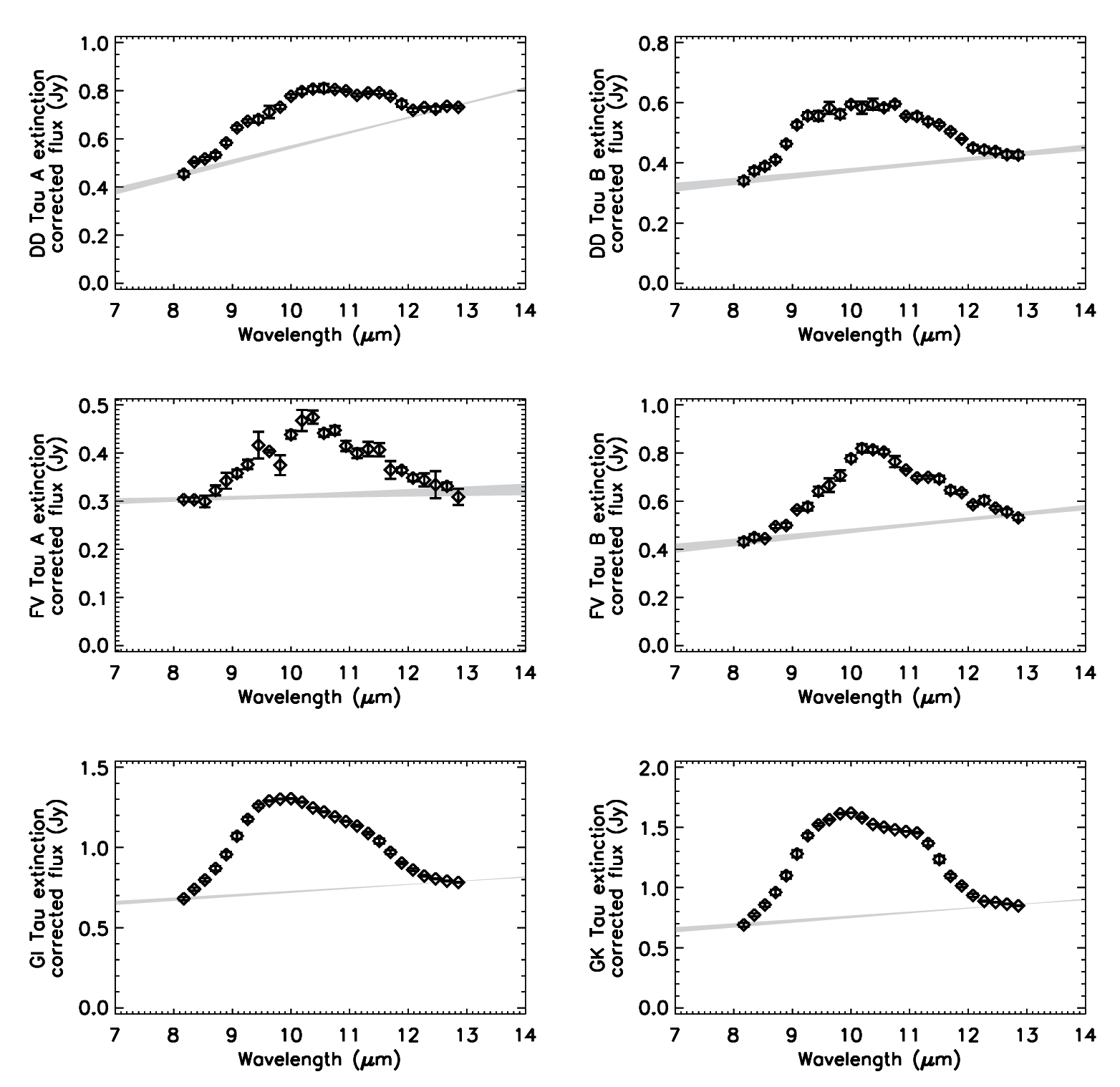}
\caption{\textit{Continued}}
\end{center}
\end{figure}

\clearpage

\begin{figure}
\begin{center}
\includegraphics[angle=0,width=6.0in]{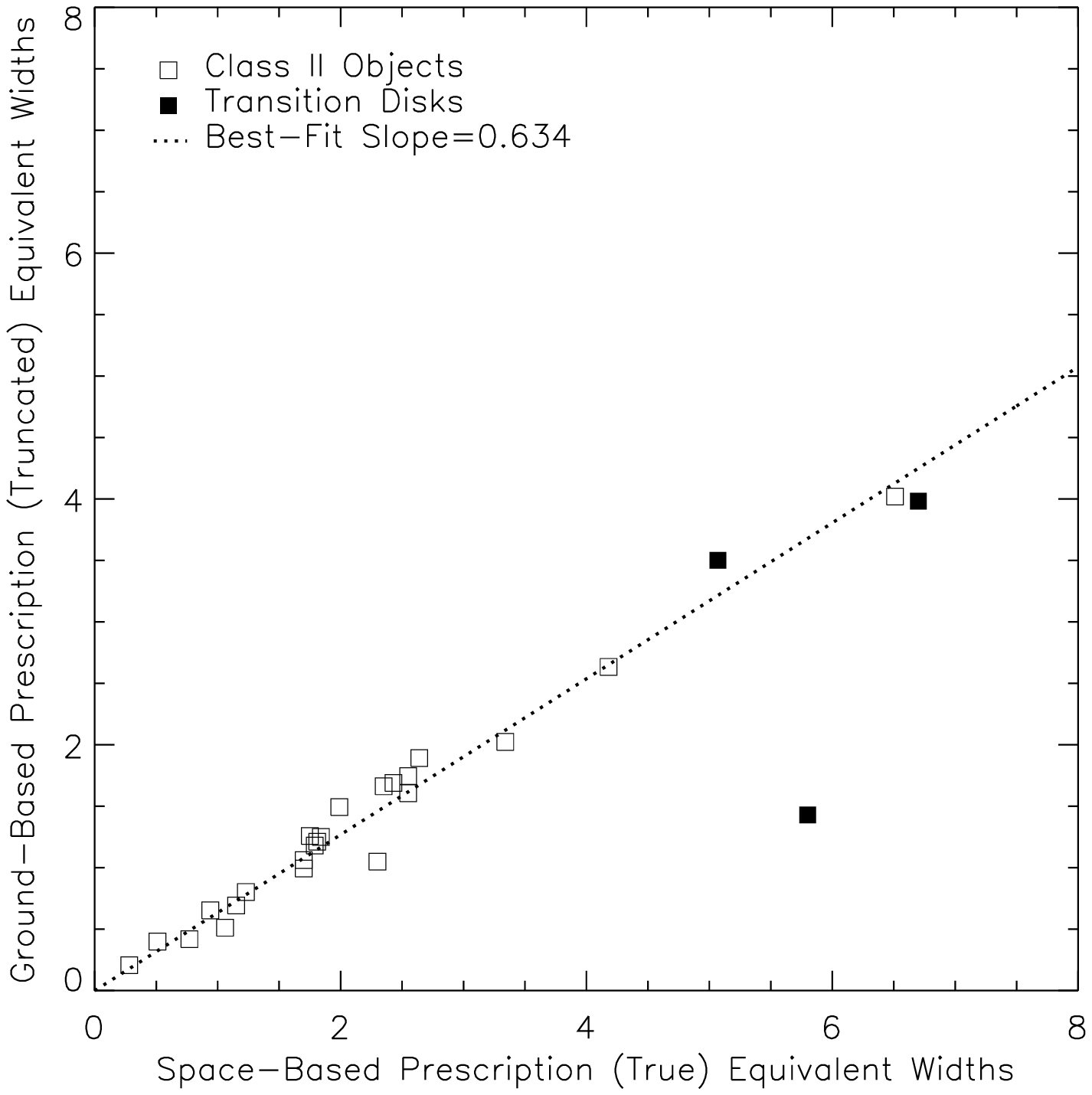}
\caption[Ground-based vs. Space-based measurements of silicate feature equivalent widths]{We take space-based equivalent width measurements from \citep{2009ApJ...703.1964F} and compare them with equivalent width using the same data but a different (i.e., ground-based) measurement prescription.  The different prescriptions produce well-correlated equivalent widths (as shown by the best-fit, dotted line), where the ground-based (truncated) equivalent widths are systematically lower by 36.6\%.  This means that ground-based (truncated) equivalent width measurements can be used as a proxy for space-based (true) equivalent-width measurements.  Note that our plot has one outlier, DM Tau, due to the large quadratic term in its continuum that is not included in our ground-based prescription.  The 3 transition disk objects all have unusually large equivalent widths.\label{Binaries EQW compare}}
\end{center}
\end{figure}

\clearpage
\begin{figure}
\begin{center}
\includegraphics[angle=90,width=6.0in]{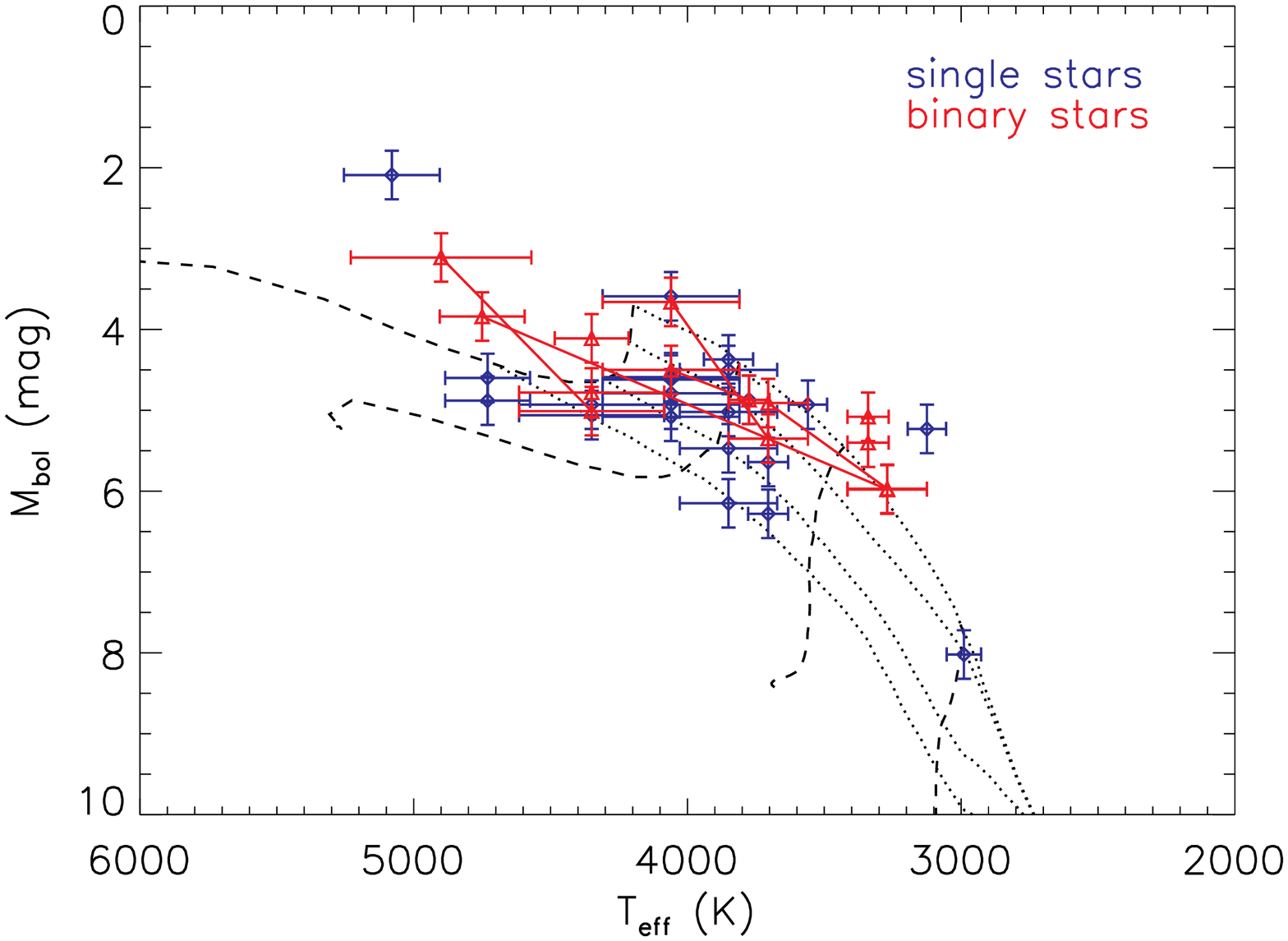}
\caption[HR diagram comparing single and binary stars]{An HR diagram showing the similar distribution of single stars (blue diamonds) and binary stars (red triangles) used in our comparison of silicate features (Section \ref{Binaries correlated?}).  Binary pairs are connected with red lines when both stars are shown.  $T_{eff}$ and $M_{bol}$ values are from \citet{2009ApJ...704..531K} when available (22 of 26 single stars and 14 of 18 stars in binaries).  Ischrones (dotted lines; 1, 2, 5 and 10 Myr) and evolutionary models (dashed lines; 0.1, 0.5, 1.0 and 1.4 $M_{\Sun}$) are from \citet{1998A&A...337..403B}.  There is a noticeable spread in the ages and masses of our sources and not all of the binary pairs fall on the same isochrones.  Some of this might be due to observations uncertainties or model errors.  In aggregate, \citet{2009ApJ...704..531K} find that Taurus binaries are more coeval than random pairs.

\label{Binaries HR}}
\end{center}
\end{figure}

\clearpage
\begin{figure}
\begin{center}
\includegraphics[angle=90,width=6.0in]{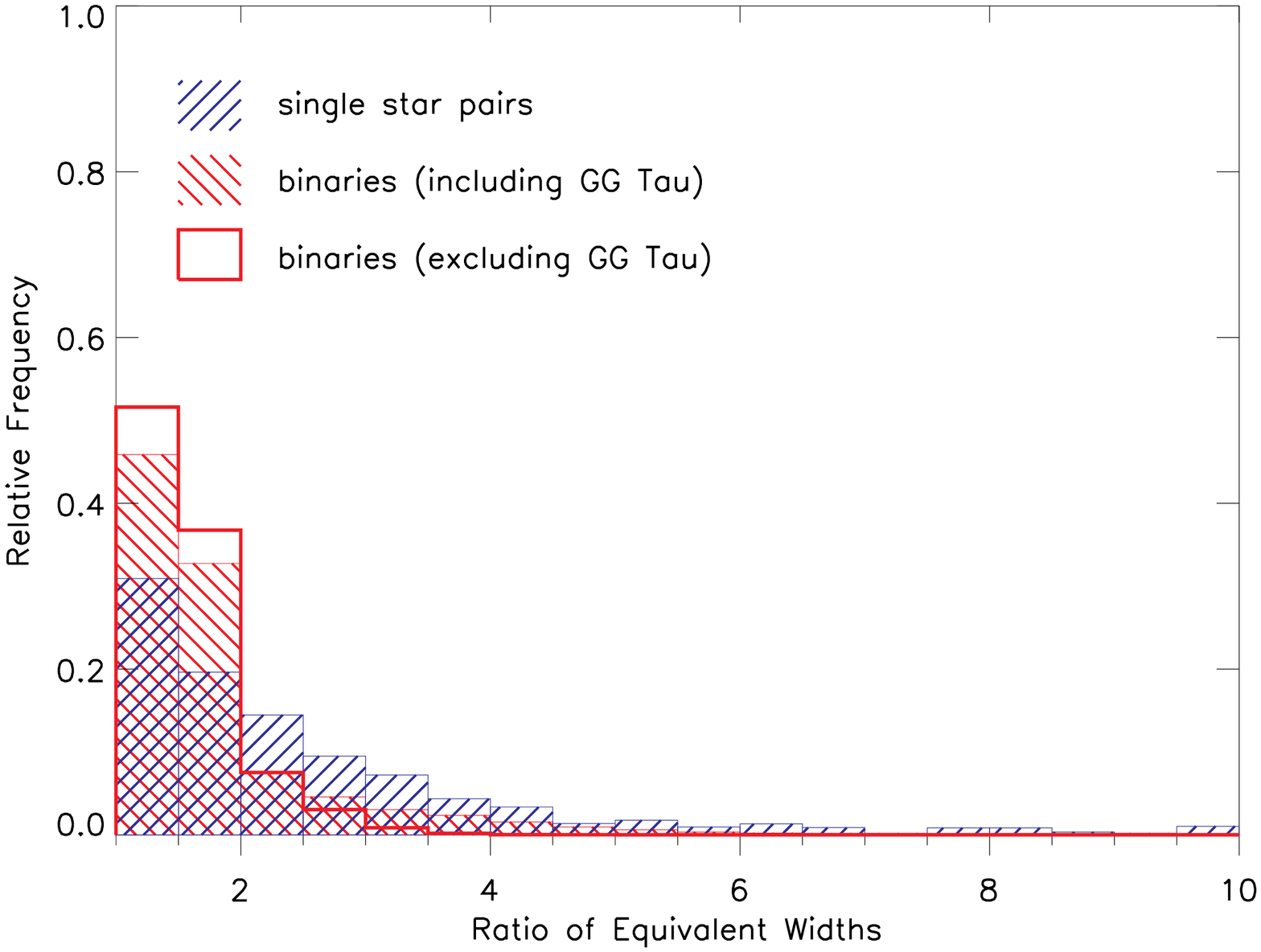}
\caption[Comparison of equivalent width ratios from our binary sample and a sample of randomly paired single stars]{A comparison of equivalent width ratios from our binary sample and a sample of randomly paired single stars.  Ratios close to 1 indicate that the silicate feature equivalent widths (which are a proxy for dust grain sizes) of a pair are similar.  When including GG Tau Aa-Ab, which might have contaminated silicate features due to accretion from circumbinary streamers, we find with 90\% confidence that the binary stars have more similar equivalent width ratios than randomly paired single stars.  When excluding GG Tau Aa-Ab, we find with 95\% confidence that the binary stars have more similar equivalent width ratios than randomly paired single stars.
\label{Binaries single_compare}}
\end{center}
\end{figure}

\clearpage
\begin{figure}
\begin{center}
\includegraphics[angle=90,width=\columnwidth]{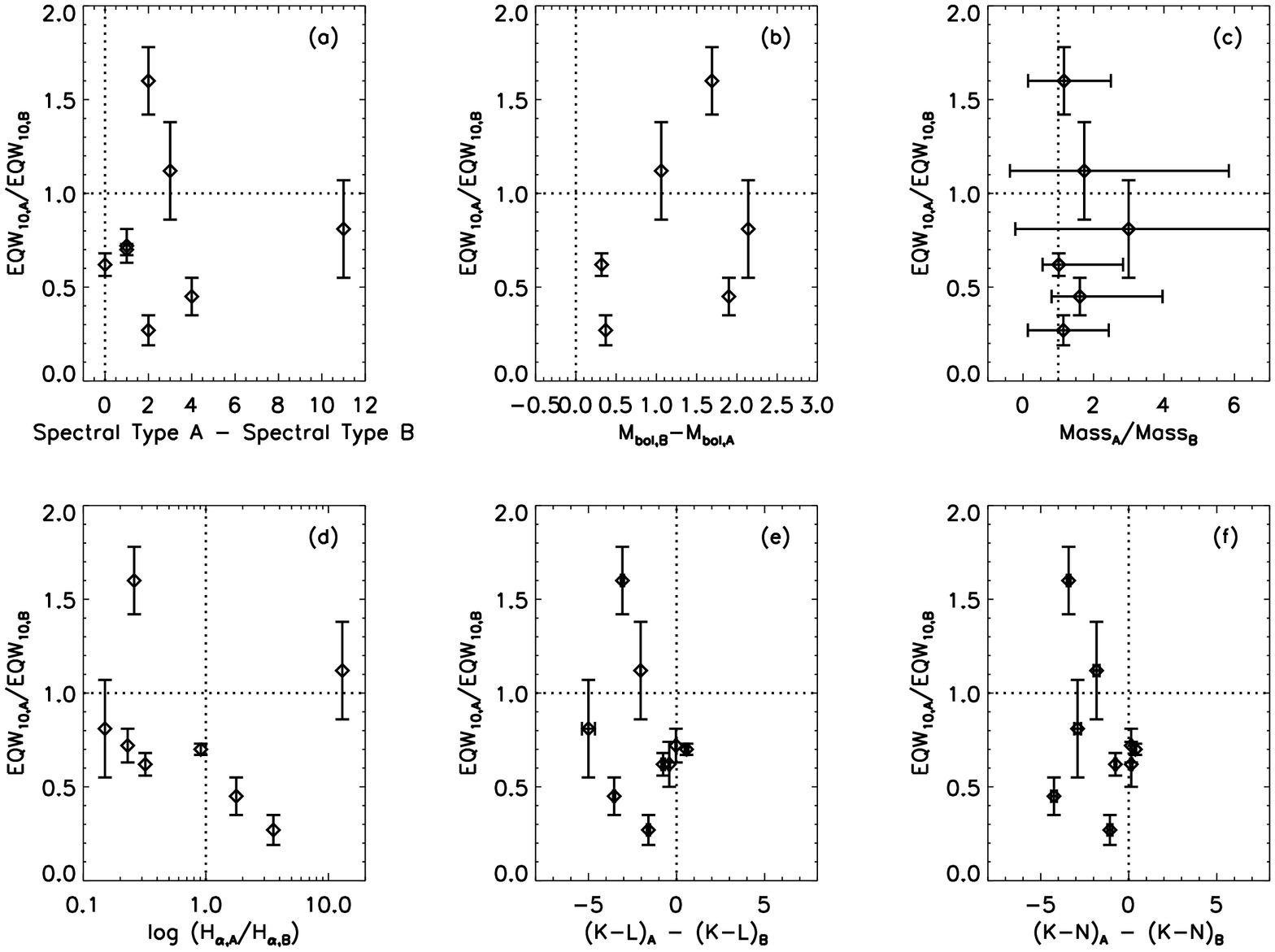}
\caption[Correlations between binary star silicate feature equivalent width ratios and a set of ancillary stellar properties and disk colors]{Correlations between binary star silicate feature equivalent width ratios and a set of ancillary stellar properties and disk colors.  The dotted lines show equal properties between binaries.  No obvious correlations are present. Frame (a) shows that in 7 out of 9 binaries, the secondary has a larger silicate feature equivalent width ($EqW_{10,A}/EqW_{10,B} < 1$).  The probability that 7 or more out of 9 binaries would have larger equivalent widths in either the earlier-type components or the later-type components is 18\%, which is low enough to warrant further study, but not low enough to prove a statistical difference.
\label{Binaries correlations}}
\end{center}
\end{figure}

\clearpage

\bibliographystyle{apj}
\bibliography{database}

\begin{thebibliography}{78}
\expandafter\ifx\csname natexlab\endcsname\relax\def\natexlab#1{#1}\fi

\bibitem[{{Apai} {et~al.}(2005){Apai}, {Pascucci}, {Bouwman}, {Natta},
  {Henning}, \& {Dullemond}}]{2005Sci...310..834A}
{Apai}, D., {Pascucci}, I., {Bouwman}, J., {Natta}, A., {Henning}, T., \&
  {Dullemond}, C.~P. 2005, Science, 310, 834

\bibitem[{{Artymowicz} \& {Lubow}(1994)}]{1994ApJ...421..651A}
{Artymowicz}, P. \& {Lubow}, S.~H. 1994, \apj, 421, 651

\bibitem[{{Artymowicz} \& {Lubow}(1996)}]{1996ApJ...467L..77A}
---. 1996, \apjl, 467, L77+

\bibitem[{{Baraffe} {et~al.}(1998){Baraffe}, {Chabrier}, {Allard}, \&
  {Hauschildt}}]{1998A&A...337..403B}
{Baraffe}, I., {Chabrier}, G., {Allard}, F., \& {Hauschildt}, P.~H. 1998, \aap,
  337, 403

\bibitem[{{Bary} {et~al.}(2009){Bary}, {Leisenring}, \&
  {Skrutskie}}]{2009ApJ...706L.168B}
{Bary}, J.~S., {Leisenring}, J.~M., \& {Skrutskie}, M.~F. 2009, \apjl, 706,
  L168

\bibitem[{{Beckwith} \& {Sargent}(1991)}]{1991ApJ...381..250B}
{Beckwith}, S.~V.~W. \& {Sargent}, A.~I. 1991, \apj, 381, 250

\bibitem[{{Brauer} {et~al.}(2008){Brauer}, {Dullemond}, \&
  {Henning}}]{2008AA...480..859B}
{Brauer}, F., {Dullemond}, C.~P., \& {Henning}, T. 2008, \aap, 480, 859

\bibitem[{{Brusa} {et~al.}(2004){Brusa}, {Miller}, {Kenworthy}, {Fisher}, \&
  {Riccardi}}]{2004SPIE.5490...23B}
{Brusa}, G., {Miller}, D.~L., {Kenworthy}, M.~A., {Fisher}, D.~L., \&
  {Riccardi}, A. 2004, in Presented at the Society of Photo-Optical
  Instrumentation Engineers (SPIE) Conference, Vol. 5490, Advancements in
  Adaptive Optics. Edited by Domenico B. Calia, Brent L. Ellerbroek, and
  Roberto Ragazzoni. Proceedings of the SPIE, Volume 5490, pp. 23-33 (2004).,
  ed. D.~{Bonaccini Calia}, B.~L. {Ellerbroek}, \& R.~{Ragazzoni}, 23--33

\bibitem[{{Carrasco-Gonz{\'a}lez} {et~al.}(2009){Carrasco-Gonz{\'a}lez},
  {Rodr{\'{\i}}guez}, {Anglada}, \& {Curiel}}]{2009ApJ...693L..86C}
{Carrasco-Gonz{\'a}lez}, C., {Rodr{\'{\i}}guez}, L.~F., {Anglada}, G., \&
  {Curiel}, S. 2009, \apjl, 693, L86

\bibitem[{{Close} {et~al.}(2003){Close}, {Biller}, {Hoffmann}, {Hinz},
  {Bieging}, {Wildi}, {Lloyd-Hart}, {Brusa}, {Fisher}, {Miller}, \&
  {Angel}}]{2003ApJ...598L..35C}
{Close}, L.~M., {Biller}, B., {Hoffmann}, W.~F., {Hinz}, P.~M., {Bieging},
  J.~H., {Wildi}, F., {Lloyd-Hart}, M., {Brusa}, G., {Fisher}, D., {Miller},
  D., \& {Angel}, R. 2003, \apjl, 598, L35

\bibitem[{{Cohen} {et~al.}(1999){Cohen}, {Walker}, {Carter}, {Hammersley},
  {Kidger}, \& {Noguchi}}]{1999AJ....117.1864C}
{Cohen}, M., {Walker}, R.~G., {Carter}, B., {Hammersley}, P., {Kidger}, M., \&
  {Noguchi}, K. 1999, \aj, 117, 1864

\bibitem[{{Dubrulle} {et~al.}(1995){Dubrulle}, {Morfill}, \&
  {Sterzik}}]{1995Icar..114..237D}
{Dubrulle}, B., {Morfill}, G., \& {Sterzik}, M. 1995, Icarus, 114, 237

\bibitem[{{Duch{\^e}ne} {et~al.}(1999){Duch{\^e}ne}, {Monin}, {Bouvier}, \&
  {M{\'e}nard}}]{Duchene_Monin_1999}
{Duch{\^e}ne}, G., {Monin}, J., {Bouvier}, J., \& {M{\'e}nard}, F. 1999, \aap,
  351, 954

\bibitem[{{Dullemond} \& {Dominik}(2005)}]{2005AA...434..971D}
{Dullemond}, C.~P. \& {Dominik}, C. 2005, \aap, 434, 971

\bibitem[{{Dutrey} {et~al.}(1996){Dutrey}, {Guilloteau}, {Duvert}, {Prato},
  {Simon}, {Schuster}, \& {Menard}}]{1996A&A...309..493D}
{Dutrey}, A., {Guilloteau}, S., {Duvert}, G., {Prato}, L., {Simon}, M.,
  {Schuster}, K., \& {Menard}, F. 1996, \aap, 309, 493

\bibitem[{{Dutrey} {et~al.}(1994){Dutrey}, {Guilloteau}, \&
  {Simon}}]{1994A&A...286..149D}
{Dutrey}, A., {Guilloteau}, S., \& {Simon}, M. 1994, \aap, 286, 149

\bibitem[{{Furlan} {et~al.}(2006){Furlan}, {Hartmann}, {Calvet}, {D'Alessio},
  {Franco-Hern{\'a}ndez}, {Forrest}, {Watson}, {Uchida}, {Sargent}, {Green},
  {Keller}, \& {Herter}}]{2006ApJS..165..568F}
{Furlan}, E., {Hartmann}, L., {Calvet}, N., {D'Alessio}, P.,
  {Franco-Hern{\'a}ndez}, R., {Forrest}, W.~J., {Watson}, D.~M., {Uchida},
  K.~I., {Sargent}, B., {Green}, J.~D., {Keller}, L.~D., \& {Herter}, T.~L.
  2006, \apjs, 165, 568

\bibitem[{{Furlan} {et~al.}(2009){Furlan}, {Watson}, {McClure}, {Manoj},
  {Espaillat}, {D'Alessio}, {Calvet}, {Kim}, {Sargent}, {Forrest}, \&
  {Hartmann}}]{2009ApJ...703.1964F}
{Furlan}, E., {Watson}, D.~M., {McClure}, M.~K., {Manoj}, P., {Espaillat}, C.,
  {D'Alessio}, P., {Calvet}, N., {Kim}, K.~H., {Sargent}, B.~A., {Forrest},
  W.~J., \& {Hartmann}, L. 2009, \apj, 703, 1964

\bibitem[{{Ghez} {et~al.}(1993){Ghez}, {Neugebauer}, \&
  {Matthews}}]{1993AJ....106.2005G}
{Ghez}, A.~M., {Neugebauer}, G., \& {Matthews}, K. 1993, \aj, 106, 2005

\bibitem[{{Glauser} {et~al.}(2009){Glauser}, {G{\"u}del}, {Watson}, {Henning},
  {Schegerer}, {Wolf}, {Audard}, \& {Baldovin-Saavedra}}]{2009A&A...508..247G}
{Glauser}, A.~M., {G{\"u}del}, M., {Watson}, D.~M., {Henning}, T., {Schegerer},
  A.~A., {Wolf}, S., {Audard}, M., \& {Baldovin-Saavedra}, C. 2009, \aap, 508,
  247

\bibitem[{{Hartigan} \& {Kenyon}(2003)}]{2003ApJ...583..334H}
{Hartigan}, P. \& {Kenyon}, S.~J. 2003, \apj, 583, 334

\bibitem[{{Hartigan} {et~al.}(1994){Hartigan}, {Strom}, \&
  {Strom}}]{Hartigan_Strom_Strom_1994}
{Hartigan}, P., {Strom}, K.~M., \& {Strom}, S.~E. 1994, \apj, 427, 961

\bibitem[{{Hinz} {et~al.}(2000){Hinz}, {Angel}, {Woolf}, {Hoffmann}, \&
  {McCarthy}}]{2000SPIE.4006..349H}
{Hinz}, P.~M., {Angel}, J.~R.~P., {Woolf}, N.~J., {Hoffmann}, W.~F., \&
  {McCarthy}, D.~W. 2000, in Society of Photo-Optical Instrumentation Engineers
  (SPIE) Conference Series, Vol. 4006, Society of Photo-Optical Instrumentation
  Engineers (SPIE) Conference Series, ed. P.~{L{\'e}na} \& A.~{Quirrenbach},
  349--353

\bibitem[{{Hoffmann} {et~al.}(1998){Hoffmann}, {Hora}, {Fazio}, {Deutsch}, \&
  {Dayal}}]{1998SPIE.3354..647H}
{Hoffmann}, W.~F., {Hora}, J.~L., {Fazio}, G.~G., {Deutsch}, L.~K., \& {Dayal},
  A. 1998, in Society of Photo-Optical Instrumentation Engineers (SPIE)
  Conference Series, Vol. 3354, Society of Photo-Optical Instrumentation
  Engineers (SPIE) Conference Series, ed. A.~M. {Fowler}, 647--658

\bibitem[{{Honda} {et~al.}(2006){Honda}, {Kataza}, {Okamoto}, {Yamashita},
  {Min}, {Miyata}, {Sako}, {Fujiyoshi}, {Sakon}, \& {Onaka}}]{Honda2006}
{Honda}, M., {Kataza}, H., {Okamoto}, Y.~K., {Yamashita}, T., {Min}, M.,
  {Miyata}, T., {Sako}, S., {Fujiyoshi}, T., {Sakon}, I., \& {Onaka}, T. 2006,
  \apj, 646, 1024

\bibitem[{{Ida} \& {Lin}(2004{\natexlab{a}})}]{2004ApJ...604..388I}
{Ida}, S. \& {Lin}, D.~N.~C. 2004{\natexlab{a}}, \apj, 604, 388

\bibitem[{{Ida} \& {Lin}(2004{\natexlab{b}})}]{2004ApJ...616..567I}
---. 2004{\natexlab{b}}, \apj, 616, 567

\bibitem[{{Jensen} {et~al.}(2007){Jensen}, {Dhital}, {Stassun}, {Patience},
  {Herbst}, {Walter}, {Simon}, \& {Basri}}]{2007AJ....134..241J}
{Jensen}, E.~L.~N., {Dhital}, S., {Stassun}, K.~G., {Patience}, J., {Herbst},
  W., {Walter}, F.~M., {Simon}, M., \& {Basri}, G. 2007, \aj, 134, 241

\bibitem[{{Jensen} {et~al.}(1996){Jensen}, {Mathieu}, \&
  {Fuller}}]{1996ApJ...458..312J}
{Jensen}, E.~L.~N., {Mathieu}, R.~D., \& {Fuller}, G.~A. 1996, \apj, 458, 312

\bibitem[{{Juh{\'a}sz} {et~al.}(2009){Juh{\'a}sz}, {Henning}, {Bouwman},
  {Dullemond}, {Pascucci}, \& {Apai}}]{2009ApJ...695.1024J}
{Juh{\'a}sz}, A., {Henning}, T., {Bouwman}, J., {Dullemond}, C.~P., {Pascucci},
  I., \& {Apai}, D. 2009, \apj, 695, 1024

\bibitem[{{Kemper} {et~al.}(2004){Kemper}, {Vriend}, \&
  {Tielens}}]{2004ApJ...609..826K}
{Kemper}, F., {Vriend}, W.~J., \& {Tielens}, A.~G.~G.~M. 2004, \apj, 609, 826

\bibitem[{{Kenyon} {et~al.}(1994){Kenyon}, {Dobrzycka}, \&
  {Hartmann}}]{1994AJ....108.1872K}
{Kenyon}, S.~J., {Dobrzycka}, D., \& {Hartmann}, L. 1994, \aj, 108, 1872

\bibitem[{{Kenyon} \& {Hartmann}(1995)}]{1995ApJS..101..117K}
{Kenyon}, S.~J. \& {Hartmann}, L. 1995, \apjs, 101, 117

\bibitem[{{Kessler-Silacci} {et~al.}(2006){Kessler-Silacci}, {Augereau},
  {Dullemond}, {Geers}, {Lahuis}, {Evans}, {van Dishoeck}, {Blake}, {Boogert},
  {Brown}, {J{\o}rgensen}, {Knez}, \& {Pontoppidan}}]{2006ApJ...639..275K}
{Kessler-Silacci}, J., {Augereau}, J., {Dullemond}, C.~P., {Geers}, V.,
  {Lahuis}, F., {Evans}, II, N.~J., {van Dishoeck}, E.~F., {Blake}, G.~A.,
  {Boogert}, A.~C.~A., {Brown}, J., {J{\o}rgensen}, J.~K., {Knez}, C., \&
  {Pontoppidan}, K.~M. 2006, \apj, 639, 275

\bibitem[{{Kessler-Silacci} {et~al.}(2007){Kessler-Silacci}, {Dullemond},
  {Augereau}, {Mer{\'{\i}}n}, {Geers}, {van Dishoeck}, {Evans}, {Blake}, \&
  {Brown}}]{2007ApJ...659..680K}
{Kessler-Silacci}, J.~E., {Dullemond}, C.~P., {Augereau}, J., {Mer{\'{\i}}n},
  B., {Geers}, V.~C., {van Dishoeck}, E.~F., {Evans}, II, N.~J., {Blake},
  G.~A., \& {Brown}, J. 2007, \apj, 659, 680

\bibitem[{{K{\"o}hler} {et~al.}(2008){K{\"o}hler}, {Ratzka}, {Herbst}, \&
  {Kasper}}]{Kohler2008}
{K{\"o}hler}, R., {Ratzka}, T., {Herbst}, T.~M., \& {Kasper}, M. 2008, \aap,
  482, 929

\bibitem[{{Kraus} \& {Hillenbrand}(2009)}]{2009ApJ...704..531K}
{Kraus}, A.~L. \& {Hillenbrand}, L.~A. 2009, \apj, 704, 531

\bibitem[{{Kruegel} \& {Siebenmorgen}(1994)}]{1994A&A...288..929K}
{Kruegel}, E. \& {Siebenmorgen}, R. 1994, \aap, 288, 929

\bibitem[{{Lissauer} \& {Stevenson}(2007)}]{2007prpl.conf..591L}
{Lissauer}, J.~J. \& {Stevenson}, D.~J. 2007, Protostars and Planets V, 591

\bibitem[{{Lloyd-Hart}(2000)}]{2000PASP..112..264L}
{Lloyd-Hart}, M. 2000, \pasp, 112, 264

\bibitem[{{Lommen} {et~al.}(2010){Lommen}, {van Dishoeck}, {Wright},
  {Maddison}, {Min}, {Wilner}, {Salter}, {van Langevelde}, {Bourke}, {van der
  Burg}, \& {Blake}}]{2010A&A...515A..77L}
{Lommen}, D.~J.~P., {van Dishoeck}, E.~F., {Wright}, C.~M., {Maddison}, S.~T.,
  {Min}, M., {Wilner}, D.~J., {Salter}, D.~M., {van Langevelde}, H.~J.,
  {Bourke}, T.~L., {van der Burg}, R.~F.~J., \& {Blake}, G.~A. 2010, \aap, 515,
  A77+

\bibitem[{{Luhman} {et~al.}(2006){Luhman}, {Whitney}, {Meade}, {Babler},
  {Indebetouw}, {Bracker}, \& {Churchwell}}]{2006ApJ...647.1180L}
{Luhman}, K.~L., {Whitney}, B.~A., {Meade}, M.~R., {Babler}, B.~L.,
  {Indebetouw}, R., {Bracker}, S., \& {Churchwell}, E.~B. 2006, \apj, 647, 1180

\bibitem[{{Mathieu} {et~al.}(1997){Mathieu}, {Stassun}, {Basri}, {Jensen},
  {Johns-Krull}, {Valenti}, \& {Hartmann}}]{1997AJ....113.1841M}
{Mathieu}, R.~D., {Stassun}, K., {Basri}, G., {Jensen}, E.~L.~N.,
  {Johns-Krull}, C.~M., {Valenti}, J.~A., \& {Hartmann}, L.~W. 1997, \aj, 113,
  1841

\bibitem[{{Mathis}(1990)}]{1990ARA&A..28...37M}
{Mathis}, J.~S. 1990, \araa, 28, 37

\bibitem[{{McClure}(2009)}]{2009ApJ...693L..81M}
{McClure}, M. 2009, \apjl, 693, L81

\bibitem[{{Meeus} {et~al.}(2001){Meeus}, {Waters}, {Bouwman}, {van den Ancker},
  {Waelkens}, \& {Malfait}}]{2001A&A...365..476M}
{Meeus}, G., {Waters}, L.~B.~F.~M., {Bouwman}, J., {van den Ancker}, M.~E.,
  {Waelkens}, C., \& {Malfait}, K. 2001, \aap, 365, 476

\bibitem[{{Monin} {et~al.}(1998){Monin}, {Menard}, \&
  {Duchene}}]{1998A&A...339..113M}
{Monin}, J., {Menard}, F., \& {Duchene}, G. 1998, \aap, 339, 113

\bibitem[{{Natta} {et~al.}(2007){Natta}, {Testi}, {Calvet}, {Henning},
  {Waters}, \& {Wilner}}]{2007prpl.conf..767N}
{Natta}, A., {Testi}, L., {Calvet}, N., {Henning}, T., {Waters}, R., \&
  {Wilner}, D. 2007, Protostars and Planets V, 767

\bibitem[{{Oliveira} {et~al.}(2011){Oliveira}, {Olofsson}, {Pontoppidan}, {van
  Dishoeck}, {Augereau}, \& {Merin}}]{2011arXiv1104.3574O}
{Oliveira}, I., {Olofsson}, J., {Pontoppidan}, K.~M., {van Dishoeck}, E.~F.,
  {Augereau}, J., \& {Merin}, B. 2011, ArXiv e-prints

\bibitem[{{Olofsson} {et~al.}(2010){Olofsson}, {Augereau}, {van Dishoeck},
  {Mer{\'{\i}}n}, {Grosso}, {M{\'e}nard}, {Blake}, \&
  {Monin}}]{2010A&A...520A..39O}
{Olofsson}, J., {Augereau}, J.-C., {van Dishoeck}, E.~F., {Mer{\'{\i}}n}, B.,
  {Grosso}, N., {M{\'e}nard}, F., {Blake}, G.~A., \& {Monin}, J.-L. 2010, \aap,
  520, A39+

\bibitem[{{Ossenkopf} \& {Henning}(1994)}]{1994A&A...291..943O}
{Ossenkopf}, V. \& {Henning}, T. 1994, \aap, 291, 943

\bibitem[{{Pascucci} {et~al.}(2008){Pascucci}, {Apai}, {Hardegree-Ullman},
  {Kim}, {Meyer}, \& {Bouwman}}]{2008ApJ...673..477P}
{Pascucci}, I., {Apai}, D., {Hardegree-Ullman}, E.~E., {Kim}, J.~S., {Meyer},
  M.~R., \& {Bouwman}, J. 2008, \apj, 673, 477

\bibitem[{{Pascucci} {et~al.}(2009){Pascucci}, {Apai}, {Luhman}, {Henning},
  {Bouwman}, {Meyer}, {Lahuis}, \& {Natta}}]{2009ApJ...696..143P}
{Pascucci}, I., {Apai}, D., {Luhman}, K., {Henning}, T., {Bouwman}, J.,
  {Meyer}, M.~R., {Lahuis}, F., \& {Natta}, A. 2009, \apj, 696, 143

\bibitem[{{Pi{\'e}tu} {et~al.}(2011){Pi{\'e}tu}, {Gueth}, {Hily-Blant},
  {Schuster}, \& {Pety}}]{2011AA...528A..81P}
{Pi{\'e}tu}, V., {Gueth}, F., {Hily-Blant}, P., {Schuster}, K., \& {Pety}, J.
  2011, \aap, 528, A81+

\bibitem[{{Przygodda} {et~al.}(2003){Przygodda}, {van Boekel},
  {{\`A}brah{\`a}m}, {Melnikov}, {Waters}, \& {Leinert}}]{2003A&A...412L..43P}
{Przygodda}, F., {van Boekel}, R., {{\`A}brah{\`a}m}, P., {Melnikov}, S.~Y.,
  {Waters}, L.~B.~F.~M., \& {Leinert}, C. 2003, \aap, 412, L43

\bibitem[{{Ratzka} {et~al.}(2009){Ratzka}, {Schegerer}, {Leinert},
  {{\'A}brah{\'a}m}, {Henning}, {Herbst}, {K{\"o}hler}, {Wolf}, \&
  {Zinnecker}}]{Ratzka}
{Ratzka}, T., {Schegerer}, A.~A., {Leinert}, C., {{\'A}brah{\'a}m}, P.,
  {Henning}, T., {Herbst}, T.~M., {K{\"o}hler}, R., {Wolf}, S., \& {Zinnecker},
  H. 2009, \aap, 502, 623

\bibitem[{{Riaz}(2009)}]{2009ApJ...701..571R}
{Riaz}, B. 2009, \apj, 701, 571

\bibitem[{{Rieke} \& {Lebofsky}(1985)}]{1985ApJ...288..618R}
{Rieke}, G.~H. \& {Lebofsky}, M.~J. 1985, \apj, 288, 618

\bibitem[{{Roddier} {et~al.}(1996){Roddier}, {Roddier}, {Northcott}, {Graves},
  \& {Jim}}]{1996ApJ...463..326R}
{Roddier}, C., {Roddier}, F., {Northcott}, M.~J., {Graves}, J.~E., \& {Jim}, K.
  1996, \apj, 463, 326

\bibitem[{{Sargent} {et~al.}(2009){Sargent}, {Forrest}, {Tayrien}, {McClure},
  {Watson}, {Sloan}, {Li}, {Manoj}, {Bohac}, {Furlan}, {Kim}, \&
  {Green}}]{2009ApJS..182..477S}
{Sargent}, B.~A., {Forrest}, W.~J., {Tayrien}, C., {McClure}, M.~K., {Watson},
  D.~M., {Sloan}, G.~C., {Li}, A., {Manoj}, P., {Bohac}, C.~J., {Furlan}, E.,
  {Kim}, K.~H., \& {Green}, J.~D. 2009, \apjs, 182, 477

\bibitem[{{Sicilia-Aguilar} {et~al.}(2007){Sicilia-Aguilar}, {Hartmann},
  {Watson}, {Bohac}, {Henning}, \& {Bouwman}}]{2007ApJ...659.1637S}
{Sicilia-Aguilar}, A., {Hartmann}, L.~W., {Watson}, D., {Bohac}, C., {Henning},
  T., \& {Bouwman}, J. 2007, \apj, 659, 1637

\bibitem[{{Simon} {et~al.}(1996){Simon}, {Holfeltz}, \&
  {Taff}}]{1996ApJ...469..890S}
{Simon}, M., {Holfeltz}, S.~T., \& {Taff}, L.~G. 1996, \apj, 469, 890

\bibitem[{{Skemer} {et~al.}(2010){Skemer}, {Close}, {Hinz}, {Hoffmann},
  {Greene}, {Males}, \& {Beck}}]{Skemer2010}
{Skemer}, A.~J., {Close}, L.~M., {Hinz}, P.~M., {Hoffmann}, W.~F., {Greene},
  T.~P., {Males}, J.~R., \& {Beck}, T.~L. 2010, \apj, 711, 1280

\bibitem[{{Skemer} {et~al.}(2008){Skemer}, {Close}, {Hinz}, {Hoffmann},
  {Kenworthy}, \& {Miller}}]{2008ApJ...676.1082S}
{Skemer}, A.~J., {Close}, L.~M., {Hinz}, P.~M., {Hoffmann}, W.~F., {Kenworthy},
  M.~A., \& {Miller}, D.~L. 2008, \apj, 676, 1082

\bibitem[{{Skemer} {et~al.}(2009){Skemer}, {Hinz}, {Hoffmann}, {Close},
  {Kendrew}, {Mathar}, {Stuik}, {Greene}, {Woodward}, \& {Kelley}}]{Skemer09}
{Skemer}, A.~J., {Hinz}, P.~M., {Hoffmann}, W.~F., {Close}, L.~M., {Kendrew},
  S., {Mathar}, R.~J., {Stuik}, R., {Greene}, T.~P., {Woodward}, C.~E., \&
  {Kelley}, M.~S. 2009, \pasp, 121, 897

\bibitem[{{Skrutskie} {et~al.}(2006){Skrutskie}, {Cutri}, {Stiening},
  {Weinberg}, {Schneider}, {Carpenter}, {Beichman}, {Capps}, {Chester},
  {Elias}, {Huchra}, {Liebert}, {Lonsdale}, {Monet}, {Price}, {Seitzer},
  {Jarrett}, {Kirkpatrick}, {Gizis}, {Howard}, {Evans}, {Fowler}, {Fullmer},
  {Hurt}, {Light}, {Kopan}, {Marsh}, {McCallon}, {Tam}, {Van Dyk}, \&
  {Wheelock}}]{2006AJ....131.1163S}
{Skrutskie}, M.~F., {Cutri}, R.~M., {Stiening}, R., {Weinberg}, M.~D.,
  {Schneider}, S., {Carpenter}, J.~M., {Beichman}, C., {Capps}, R., {Chester},
  T., {Elias}, J., {Huchra}, J., {Liebert}, J., {Lonsdale}, C., {Monet}, D.~G.,
  {Price}, S., {Seitzer}, P., {Jarrett}, T., {Kirkpatrick}, J.~D., {Gizis},
  J.~E., {Howard}, E., {Evans}, T., {Fowler}, J., {Fullmer}, L., {Hurt}, R.,
  {Light}, R., {Kopan}, E.~L., {Marsh}, K.~A., {McCallon}, H.~L., {Tam}, R.,
  {Van Dyk}, S., \& {Wheelock}, S. 2006, \aj, 131, 1163

\bibitem[{{Stapelfeldt} {et~al.}(1998){Stapelfeldt}, {Krist}, {Menard},
  {Bouvier}, {Padgett}, \& {Burrows}}]{1998ApJ...502L..65S}
{Stapelfeldt}, K.~R., {Krist}, J.~E., {Menard}, F., {Bouvier}, J., {Padgett},
  D.~L., \& {Burrows}, C.~J. 1998, \apjl, 502, L65+

\bibitem[{{van Boekel} {et~al.}(2010){van Boekel}, {Juh{\'a}sz}, {Henning},
  {K{\"o}hler}, {Ratzka}, {Herbst}, {Bouwman}, \& {Kley}}]{2010AA...517A..16V}
{van Boekel}, R., {Juh{\'a}sz}, A., {Henning}, T., {K{\"o}hler}, R., {Ratzka},
  T., {Herbst}, T., {Bouwman}, J., \& {Kley}, W. 2010, \aap, 517, A16+

\bibitem[{{van Boekel} {et~al.}(2005){van Boekel}, {Min}, {Waters}, {de Koter},
  {Dominik}, {van den Ancker}, \& {Bouwman}}]{2005A&A...437..189V}
{van Boekel}, R., {Min}, M., {Waters}, L.~B.~F.~M., {de Koter}, A., {Dominik},
  C., {van den Ancker}, M.~E., \& {Bouwman}, J. 2005, \aap, 437, 189

\bibitem[{{van Boekel} {et~al.}(2003){van Boekel}, {Waters}, {Dominik},
  {Bouwman}, {de Koter}, {Dullemond}, \& {Paresce}}]{2003A&A...400L..21V}
{van Boekel}, R., {Waters}, L.~B.~F.~M., {Dominik}, C., {Bouwman}, J., {de
  Koter}, A., {Dullemond}, C.~P., \& {Paresce}, F. 2003, \aap, 400, L21

\bibitem[{{Watson}(2009)}]{2009ASPC..414...77W}
{Watson}, D. 2009, in Astronomical Society of the Pacific Conference Series,
  Vol. 414, Cosmic Dust - Near and Far, ed. {T.~Henning, E.~Gr{\"u}n, \&
  J.~Steinacker}, 77--+

\bibitem[{{Watson} {et~al.}(2009){Watson}, {Leisenring}, {Furlan}, {Bohac},
  {Sargent}, {Forrest}, {Calvet}, {Hartmann}, {Nordhaus}, {Green}, {Kim},
  {Sloan}, {Chen}, {Keller}, {d'Alessio}, {Najita}, {Uchida}, \&
  {Houck}}]{Watson2009}
{Watson}, D.~M., {Leisenring}, J.~M., {Furlan}, E., {Bohac}, C.~J., {Sargent},
  B., {Forrest}, W.~J., {Calvet}, N., {Hartmann}, L., {Nordhaus}, J.~T.,
  {Green}, J.~D., {Kim}, K.~H., {Sloan}, G.~C., {Chen}, C.~H., {Keller}, L.~D.,
  {d'Alessio}, P., {Najita}, J., {Uchida}, K.~I., \& {Houck}, J.~R. 2009,
  \apjs, 180, 84

\bibitem[{{Weidenschilling}(1977)}]{1977MNRAS.180...57W}
{Weidenschilling}, S.~J. 1977, \mnras, 180, 57

\bibitem[{{White} \& {Basri}(2003)}]{2003ApJ...582.1109W}
{White}, R.~J. \& {Basri}, G. 2003, \apj, 582, 1109

\bibitem[{{White} \& {Ghez}(2001)}]{White_Ghez_2001}
{White}, R.~J. \& {Ghez}, A.~M. 2001, \apj, 556, 265

\bibitem[{{Wildi} {et~al.}(2003){Wildi}, {Brusa}, {Lloyd-Hart}, {Close}, \&
  {Riccardi}}]{2003SPIE.5169...17W}
{Wildi}, F.~P., {Brusa}, G., {Lloyd-Hart}, M., {Close}, L.~M., \& {Riccardi},
  A. 2003, in Society of Photo-Optical Instrumentation Engineers (SPIE)
  Conference Series, Vol. 5169, Society of Photo-Optical Instrumentation
  Engineers (SPIE) Conference Series, ed. R.~K. {Tyson} \& M.~{Lloyd-Hart},
  17--25

\bibitem[{{Wooden} {et~al.}(2000){Wooden}, {Bell}, {Harker}, \&
  {Woodward}}]{2000AAS...197.4707W}
{Wooden}, D.~H., {Bell}, K.~R., {Harker}, D.~E., \& {Woodward}, C.~E. 2000, in
  Bulletin of the American Astronomical Society, Vol.~32, Bulletin of the
  American Astronomical Society, 1482--+

\bibitem[{{Zsom} {et~al.}(2011){Zsom}, {S{\'a}ndor}, \&
  {Dullemond}}]{2011A&A...527A..10Z}
{Zsom}, A., {S{\'a}ndor}, Z., \& {Dullemond}, C.~P. 2011, \aap, 527, A10+

\end{thebibliography}

\end{document}